\documentclass[a4paper,12pt]{article}

\usepackage{natbib}
\usepackage{graphicx,amsmath}
\newcommand{\ain}{{a_{\rm in}}}
\newcommand{\aout}{{a_{\rm out}}}

\newcommand{\dd}{\delta}

\newcommand{\elie}{{\mathbf E}}

\newcommand{\cotan}{\mathrm{cotan} \;}

\newcommand{\psirect}{\psi^{\rm rect.} }

\newcommand{\psir}{\psi^{\rm annulus} }
\newcommand{\psib}{\psi^{\rm float} }
\newcommand{\psicd}{\psi^{\rm disc} }
\newcommand{\psipod}{\psi^{\rm sector}}% \, of \, disc} }
\newcommand{\gr}{g_R }
\newcommand{\gt}{g_\theta }

\newcommand{\psicell}{\psi^{\rm cell} }
\newcommand{\sign}{\,{\rm Sign} }
\newcommand{\asinh}{\,{\rm asinh \;} }

\newcommand{\elik}{{\mathbf K}}
\newcommand{\elipi}{{\mathbf \Pi}}        
\newcommand{\pdz}{\partial_Z}
\newcommand{\pdr}{\partial_R}       
\newcommand{\pdtheta}{\partial_\theta}       

\newcommand{\mnras}{Mon. Not. of the Royal Astr. Society}
\newcommand{\aap}{Astron. \& Astrophys.}

\newcommand{\apj}{Astrophys. J.}
\newcommand{\aj}{Astron. J.}

\def\atan{{\rm arctan \;}}

\begin{document}

\title{A key-formula to compute the gravitational potential of inhomogeneous discs in cylindrical coordinates}

\author{Jean-Marc Hur\'e$^{1,2}$\\\\\small $^1$Univ. Bordeaux, LAB, UMR 5804, F-33270, Floirac, France\\\small $^2$CNRS, LAB, UMR 5804, F-33270, Floirac, France\\\\
\small email \tt{jean-marc.hure@obs.u-bordeaux1.fr}\\\\
To appear in {\it Celestial Mechanics and Dynamical Astronomy}
}
%\institute{Univ. Bordeaux, LAB, UMR 5804, F-33270, Floirac, France \and CNRS, LAB, UMR 5804, F-33270, Floirac, France\\
%\email{jean-marc.hure@obs.u-bordeaux1.fr}
%}


\maketitle

\begin{abstract}
We have established the exact expression for the gravitational potential of a homogeneous polar cell | an elementary pattern used in hydrodynamical simulations of gravitating discs. This formula, which is a closed-form, works for any opening angle and radial extension of the cell. It is valid at any point in space, i.e. in the plane of the distribution (inside and outside) as well as off-plane, thereby generalizing the results reported by Durand (1953) for the circular disc. The three components of the gravitational acceleration are given. The mathematical demonstration proceeds from the {\it incomplete version of Durand's formula} for the potential (based on complete elliptic integrals). We determine first the potential due to the circular sector (i.e. a pie-slice sheet), and then deduce that of the polar cell (from convenient radial scaling and subtraction). As a by-product, we generate an integral theorem stating that {\it "the angular average of the potential of any circular sector along its tangent circle is $\frac{2}{\pi}$ times the value at the corner''}. A few examples are presented. For numerical resolutions and cell shapes commonly used in disc simulations, we quantify the importance of curvature effects by performing a direct comparison between the potential of the polar cell and that of the Cartesian (i.e. rectangular) cell having the {\it same mass}. Edge values are found to deviate roughly like $2 \times 10^{-3} \times N/256$ in relative ($N$ is the number of grid points in the radial direction), while the agreement is typically four orders of magnitude better for values at the cell's center. We also produce a reliable approximation for the potential, valid in the cell's plane,  inside and close to the cell. Its remarkable accuracy, about $5 \times 10^{-4} \times N/256$ in relative, is sufficient to estimate the cell's self-acceleration. 
\end{abstract}

%\PACS{95.30.Sf, 41.20.Cv, 02.60.-x, 04.40.-b}
%\keywords{Gravity \and Disc \and Analytical methods \and Numerical methods \and Elliptic integrals}

%\newpage

\section{Introduction}

The determination of the potential and forces due to various distributions of matter is an important part of theoretical and numerical geophysics and especially astrophysics. Applications concern for instance the dynamics of solid particles (satellites, asteroids, circumstellar material, planetary rings) orbiting dense objects, the internal structure of self-gravitating systems like planets, rotating stars, galaxies and massive discs, moments of inertia and geodetic problems \cite[e.g.][]{binneytremaine87,as94,subrkaras05,nelson06,bm08,vl09,schulz09}. In general, symmetry in shape or in mass distribution offers a simplified mathematical framework. However, this is not always sufficient to enable a full analytical calculus. Other assumptions are therefore sometimes invoked, like the reduction of the number of dimensions. The circular disc as a flat, two dimensional system is of great interest in astrophysics. Several cases have been studied, differing mainly by the size of the system (finite or not), and by the surface density profile $\Sigma$, generally a function of the radius $a$ and polar angle, the most basic being probably the homogeneous disc \citep{durand64,lassblitzer83,conway00,fukushima10,tresaco11}. Non-homogeneous discs have been widely considered in the context of galactic dynamics. There are axially symmetrical discs, analysed for instance by \cite{mestel63} for $\Sigma \propto 1/a$, by \cite{casertano83} for $\Sigma \propto e^{-a}$, by \cite{schulz09,schulz11} for $\Sigma \propto (1-a)^{n/2}$ ($n$ integer), by \cite{hpelatp07} and \cite{bradamilgrom95} for $\Sigma \propto a^s$ (with $s$ real). Potential/surface density pairs have also been produced for fully inhomogeneous discs \cite{ka76,qian93,qian92,evanscollett93} \citep[see also][]{binneytremaine87}.

In models and especially in numerical simulations working in polar and cylindrical coordinates, circular discs are naturally discretized into a large collection of massive polar cells, which individually generate in space an attractive force. This enables to introduce any kind of inhomogeneity from {\it individual homogeneous cells}. To our knowledge, the general formula for the gravitational potential of a polar cell | leading to these forces | is not known yet, probably because it is not really easy to derive from direct integral calculus. \cite{hph09} have addressed the problem in the special case where the potential and field is required along the symmetry axis of the homogeneous cell, which greatly simplifies the mathematical derivation. Here, we generalize this result and report the exact expressions valid inside the distribution as well as outside, including the complicated off-plane case. These expressions work for any cell shape (radially or angularly elongated). These new results are therefore appropriate to estimate correctly the potential and internal forces of self-gravitating discs as continuous (i.e. non-particle) systems.

The paper is organized as follows. In Section \ref{sec:disc}, we recall the formula for the potential and acceleration of a flat and homogeneous circular disc derived by \cite{durand64} in the context of electrostatics. In Section \ref{sec:strategy}, we write the integral expression for the potential of a circular sector (and polar cell) and explain our strategy to perform this integration analytically from Durand's formula. In Section \ref{sec:piece}, we deduce the generic formula enabling to treat the circular sector and polar cell as well. We give in Section \ref{sec:acceleration} the expressions enabling to construct the $3$ components of the gravitational acceleration. In Section \ref{sec:remarks}, we address a few remarks about technical aspects (computation in the cell's plane, potential in the polar axis, long-range behavior). The formula for the potential is graphically illustrated in Section \ref{sec:examples}. In Section \ref{sec:theorem}, we generate a theorem about the angular average of the potential in the cell's plane, deduced from periodicity. Section \ref{sec:curvature} is devoted to a comparison of the potential of the polar cell with its Cartesian analog, in conditions typical of current numerical simulation of hydrodynamical discs where polar cells have almost rectangular shapes. This helps to decide when and where curvature effects can be neglected. Finally, in Section \ref{sec:ax}, we establish a very good approximation for the potential, by using the expansion of the elliptic integral of the first kind around the singularity by \cite{vdv69}. A few concluding remarks follow. Several appendices contain definitions and demonstrations.

\section{Potential of a flat, circular disc}
\label{sec:disc}

\begin{figure}
\centering
\includegraphics[width=11cm,bb=0 0 582 583,clip==]{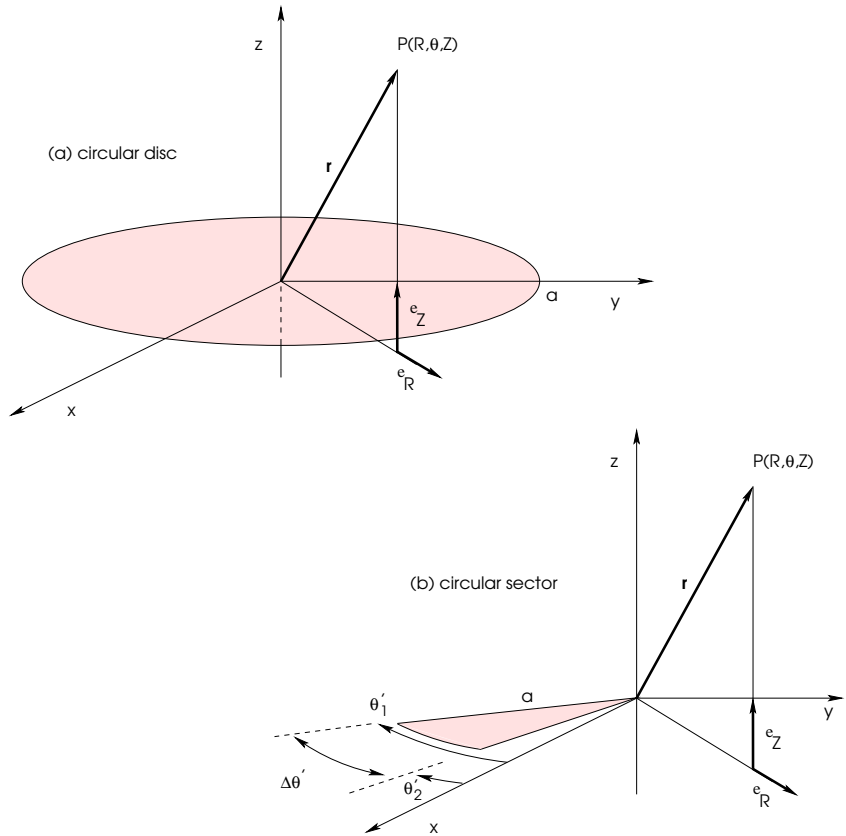}
\caption{Configuration for a circular disc with radius $a$ ({\it top panel},) and for a circular sector with radius $a$, bounding angles $\theta_1'$ and $\theta_2'$ (negative here), and opening angle $\Delta \theta'$ ({\it bottom panel}).}
\label{fig:disc.eps}
\end{figure}

The derivation of the expression for the potential of a homogeneous and flat, circular disc, with radius $a$ as shown in Fig. \ref{fig:disc.eps}a is found in \cite{durand64} in his remarkable textbook devoted to Electrostatics \citep[for the astrophysical context, see][]{lassblitzer83,conway00,tresaco11}. Up to a factor $-G \Sigma$ where $G$ is the constant of gravitation and $\Sigma$ is the surface density of the disc, the expression reads, retaining Durand's notation:
\begin{flalign}
\label{eq:psi_as}
\psicd(\vec{r};a)=2 \left[ -\pi|\zeta| \epsilon' + \dd \elie(k) + \frac{a^2-R^2}{\dd} \elik(k) % \right.\\ \nonumber
%& \qquad \qquad \qquad  \left.
 + \frac{\zeta^2}{\dd}  \frac{a-R}{a+R} \elipi(m^2,k) \right],
\end{flalign}
where $R$ and $Z$ are the cylindrical coordinates of the observation point P$(\vec{r})$  (i.e. $\vec{r} = R \vec{e}_R+ Z\vec{e}_Z$), $\elie$, $\elik$ and $ \elipi$ are the complete elliptic integrals of the first, second and third kinds respectively (see the Appendix \ref{app:ei} for a definition),
\begin{equation}
\zeta=Z-z,
\end{equation}
where $z$ is the altitude of the disc,
\begin{equation}
\label{eq:r1}
\dd = \sqrt{(a+R)^2+\zeta^2},
\end{equation}
\begin{equation}
\label{eq:kmod}
k = \frac{2\sqrt{aR}}{\dd}
\end{equation}
is the modulus ($0 \le k \le 1$),
\begin{equation}
\label{eq:mmod}
m = \frac{2\sqrt{aR}}{(a+R)}
\end{equation}
is the characteristic or parameter ($0 \le k \le m \le 1$), and
\begin{equation}
\epsilon' =
\begin{cases}
1, \quad  {\rm if } \; R<a,\\
\frac{1}{2}, \quad   {\rm if } \; R=a,\\
0, \quad  {\rm if } \; R>a,
\end{cases}
\end{equation}
which is a Heaviside-type function, i.e. $\epsilon'+$H$(R-a)=1$. he potential is finite and continuous everywhere, including at the edge (i.e. at $R=a$ and $Z=z$). It is possible to express the $\elipi$-function in terms of Heuman's lambda function \citep{conway00}.

We can easily calculate the potential everywhere in space of {\it an annulus}, simply by considering Eq.(\ref{eq:psi_as}) with two different values of $a$. For instance, if $\ain$ denotes the inner edge and $\aout$ is the outer edge of the annulus, then we have:
\begin{equation}
\psir(\vec{r};\ain,\aout)=\psicd(\vec{r};\aout)-\psicd(\vec{r};\ain).
\end{equation}
Equation (\ref{eq:psi_as}) is also the starting point to determine the potential of any axially symmetrical system with non-zero thickness, by performing an integration along the $z$-direction. This is usually achieved through a numerical process. Finally, from Eq.(\ref{eq:psi_as}), we can get the gravitational acceleration $\vec{g}^{\rm \, disc}=-\nabla \psicd$ by differentiation (see appendix \ref{app:g}), which is an important quantity to study the dynamics of test particles. Note however that $\vec{g}^{\rm \, disc}$ logarithmically diverges at the edge (like for any homogeneous, bidimensional distribution).

\section{Relaxing the axial symmetry: strategy of calculus}
\label{sec:strategy}

\begin{figure}
\centering
\includegraphics[width=11cm,bb=0 0 596 601,clip==]{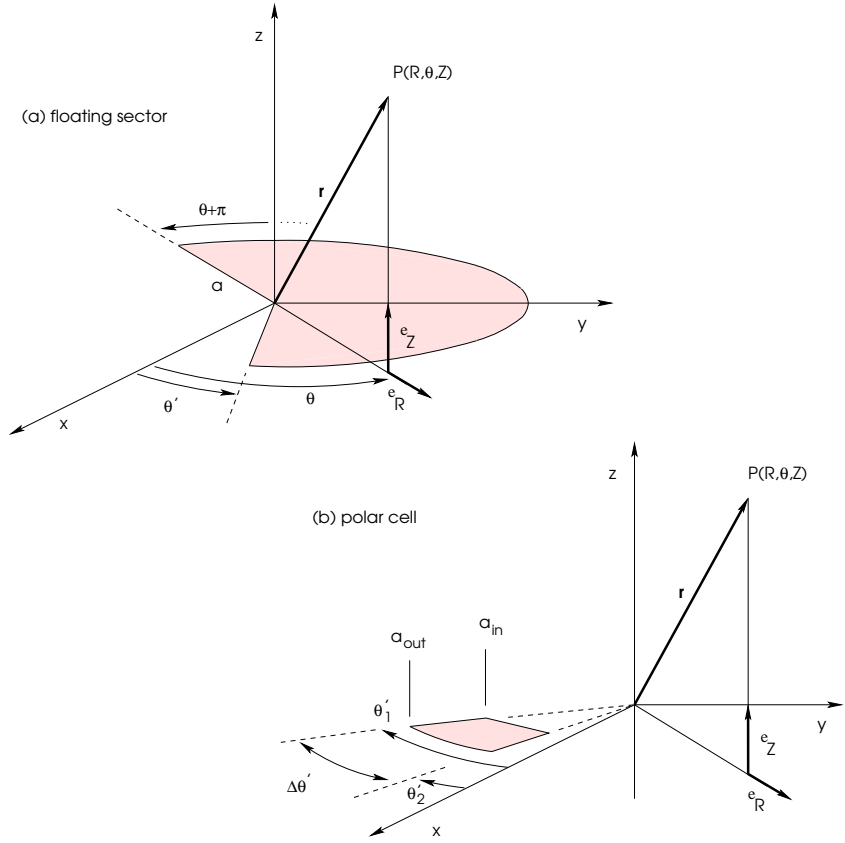}
\caption{A ``floating'' circular sector ({\it top panel}) is a particular circular sector with radius $a$ and bounding angles $\theta_1'=\theta'$ and $\theta_2'=\theta+\pi$ (this second angle depends on the observation point, and so the floating sector is not very physical). Its opening angle is $\Delta \theta' =\theta+\pi -\theta'$. Configuration for a polar cell  ({\it bottom panel}) with inner radius $\ain$, outer radius $\aout$, and bounding angles $\theta_1'$ and $\theta_2'$.}
\label{fig:disc3.eps}
\end{figure}

We now consider a homogeneous {\it circular sector} as the material domain bounded by two polar angles\footnote{As usual, polar angles are measured positively from $\vec{e}_x$. \label{note:pangle}} $\theta_1'$ and $\theta_2' \ge \theta_1'$, the polar axis $R=0$, and the circular arc with radius $a$. The corresponding configuration is shown in Fig. \ref{fig:disc.eps}b. The opening angle is $\Delta \theta' = \theta_2'- \theta_1'$, and the mass (per unit surface density) is:
\begin{equation}
M=\frac{1}{2} a^2 \Delta \theta'.
\end{equation}
In the following, we assume, without loss of generality, that $\theta_1'$ and $\theta_2'$ are in the range $[-\pi,\pi]$. This enables us to describe any circular sector, ranging from the line (the asymptotic case where $\theta_2' \rightarrow \theta_1'$) to the circular disc (the case with $\theta_2'=-\theta_1'=\pi$). To get the associated potential in space, it is necessary to go back to the integral expression which is\footnote{To lighten the notation, we use here the same variable $a$ for the integration variable and for the integral upper bound.} \citep[e.g.][]{pz05}:
\begin{flalign}
\label{eq:psipod}
\psipod(\vec{r};a,\beta_1,\beta_2) = \int_0^a{\sqrt{\frac{a}{R}} k \left[ F(\beta_1,k)-F(\beta_2,k) \right] da},
\end{flalign}
where the amplitude $\beta$ of the incomplete elliptic integral of the first kind $F$  (see the Appendix \ref{app:ei} for a definition) is defined by:
\begin{equation}
\pi - (\theta'-\theta)=2\beta,
\end{equation}
and $\theta$ denotes the polar angle of the observation point P (see note \ref{note:pangle}). We then have $2\beta_1=\pi - (\theta_1'-\theta)$ and $2\beta_2=\pi - (\theta_2'-\theta)$. Since $\beta_2 < \beta_1$, the above integral is always positive. With the above definition, we see that the quantity:
\begin{equation}
\label{eq:psifloat}
\int_0^a{\sqrt{\frac{a}{R}} k F(\beta,k)da} \equiv \psib(\vec{r};a,\beta)
\end{equation}
represents the potential of a circular sector defined by the two bounding angles $\theta_1'=\theta'$ and $\theta_2'=\theta+\pi$ (i.e. $\beta=0$). Such an object is not very physical since the second bounding angle depends on the polar angle $\theta$ of the observation point P (see below for $\theta=\theta'$); it is introduced for convenience as it plays a central role in the calculations throughout. We call such a circular sector {\it a floating circular sector}. It is displayed in Fig. \ref{fig:disc3.eps}a (in the case where $\beta > \frac{\pi}{2}$). The potential of a circular sector is then given by the difference of the potential of two floating sectors, namely:
\begin{equation}
\label{eq:psipodbasic}
\psipod(\vec{r};a,\beta_1,\beta_2) = \psib(\vec{r};a,\beta_1)  -\psib(\vec{r};a,\beta_2).
\end{equation}

We note that for $\theta_2'=-\theta_1'=\pi=$ (i.e. the circular disc), we have $2\beta_1=2\pi-\theta$ and $2\beta_2=\theta$. In these conditions, we get:
\begin{flalign}
\label{eq:fbetalarge}
 F(\beta_1,k) -F(\beta_2,k) & = F(\pi+\frac{\theta}{2},k) - F(\frac{\theta}{2},k)\\ \nonumber
& = F(\pi,k) + F(\frac{\theta}{2},k) - F(\frac{\theta}{2},k)\\ \nonumber
& = 2\elik(k)
\end{flalign}
and so Eq.(\ref{eq:psipod}) takes the well-known form:
\begin{equation}
\psicd(\vec{r};a)= 2\int_0^a{\sqrt{\frac{a}{R}} k \elik(k)da},
\end{equation}
whose closed-form is therefore Eq.(\ref{eq:psi_as}). In the special case where $\beta = \frac{\pi}{2}$ (i.e. $\theta=\theta'$), the floating sector is exactly a half-disc, and the observation point P is somewhere along the diameter. As a consequence, $\psib$ is half the potential of the circular disc, i.e. $\frac{1}{2}\psicd(\vec{r};a)$. For amplitudes $\beta \ne \frac{\pi}{2}$, the determination of Eq.(\ref{eq:psifloat}) by analytical means is not trivial. We can distinguish two cases:
\begin{itemize} 
\item within the plane of the distribution where $\zeta=0$, we have $k=m$. This is the most simple case: the result can be found either by a change of integration variable ($da \rightarrow dm$), or by using a change of modulus of the $F$-function (i.e. $m \rightarrow a/R\le 1$ or $m \rightarrow R/a \le 1$). This is the method used in \cite{hph09} to determine the potential along the symmetry axis of the polar cell.
\item when the observation point is off the plane of the sheet (i.e. $\zeta \ne 0$), things are more tricky (the goal of the paper). We have not been able to solve this general problem by direct integration but by exploring the {\it incomplete version} of Durand's formula. It turns out that this approach was successful.
\end{itemize}

To get the {\it incomplete version} of Durand's formula, we simply change the complete elliptic integrals into incomplete ones. For each term in Eq.(\ref{eq:psi_as}), the correspondence is then:
\begin{equation}
\label{eq:3der}
\begin{cases}
 \dd \elie(k) \quad \rightarrow \quad \dd E(\beta,k), \\
 \frac{a^2-R^2}{\dd}\elik(k) \quad \rightarrow \quad \frac{a^2-R^2}{\dd}F(\beta,k), \\
\frac{a-R}{a+R}\frac{\zeta^2}{\dd} \elipi(m^2,k) \quad \rightarrow \quad \frac{a-R}{a+R}\frac{\zeta^2}{\dd} \Pi(\beta,m^2,k),
\end{cases}
\end{equation}
where $E$ and $\Pi$ are the incomplete elliptic integrals of the second and third kinds respectively (see Appendix \ref{app:ei} for a definition). We have then calculated the partial derivatives of each term with respect to the radius $a$, aiming at the generation of the right integrand $\sqrt{\frac{a}{R}} k F(\beta,k)$ appearing in Eq.(\ref{eq:psifloat}). This process requires the calculation of the partial derivatives of the incomplete elliptic integrals with respect to $k$ and/or $m$ (see Appendix \ref{app:dei}), as well as the derivatives of the modulus $k$ and parameter $m$ with respect to $a$, namely:
\begin{equation}
\frac{1}{k}\partial_a k =  \frac{\zeta^2+R^2-a^2}{2a\dd^2},
\label{eq:partialka}
\end{equation}
and
\begin{equation}
\frac{1}{m}\partial_a m = \frac{R-a}{2a(a+R)}.
\label{eq:partialma}
\end{equation}
The derivatives of the three terms in the right-hand-side of Eq.(\ref{eq:3der}) are fully expanded in Appendix \ref{app:3der}. By summing the resulting formulae, we find that i) the incomplete elliptic integrals of the second and the third kinds disappear, and ii) the expected integrand is indeed present. We actually obtain the following relation:
\begin{flalign}
\label{eq:almostfinal}
 & \partial_a  \left[ \dd E(\beta,k) + \frac{a^2-R^2}{\dd}F(\beta,k)  +\frac{a-R}{a+R}\frac{\zeta^2}{\dd} \Pi(\beta,m^2,k)  \right] \\ \nonumber
& \qquad \qquad = \sqrt{\frac{a}{R}} k F(\beta,k)+  \frac{m^2 \dd}{2a} \sin \beta \cos \beta  \frac{\sqrt{1-k^2 \sin^2 \beta}}{1-m^2 \sin^2 \beta}.
\nonumber
\end{flalign}
and we recognize the expected integrand $\sqrt{\frac{a}{R}} k F(\beta,k)$. The point is that {\it the second asymmetric term can be written in the form of a partial derivative} with respect to $a$ too. This was a priori not guaranteed, but there is indeed a closed-form. The demonstration is given in Appendix \ref{app:intasymetric}. For any angle $\beta \le \frac{\pi}{2}$, we have the general relationship:
\begin{flalign}
\label{eq:daasym}
\frac{m^2 \dd}{2a} \frac{\sin \beta \cos \beta \sqrt{1-k^2 \sin^2 \beta}}{(1-m^2 \sin^2 \beta)} & = \partial_a \left\{ R \sin 2 \beta 
\asinh \frac{a+R \cos 2 \beta}{\sqrt{\zeta^2+R^2 \sin^2 2 \beta}} \right. \\\nonumber
& \qquad \left.  + \zeta \atan \left[ \frac{\zeta(a+R\cos 2 \beta)}{R \sin 2\beta \sqrt{\dd^2 - 4aR \sin^2 \beta}} \right]\right\},
\end{flalign}
and so we deduce the indefinite integral:
\begin{flalign}
\label{eq:integkF}
\int{\sqrt{\frac{a}{R}} k F(\beta,k)da}  =& \dd E(\beta,k) + \frac{a^2-R^2}{\dd}F(\beta,k) + \frac{a-R}{a+R}\frac{\zeta^2}{\dd} \Pi(\beta,m^2,k)\\ \nonumber
& -   R \sin 2\beta   \asinh \frac{a+R \cos 2 \beta}{\sqrt{\zeta^2+R^2 \sin^2 2 \beta}}   \\ \nonumber
& - \zeta \atan \left[ \frac{\zeta(a+R\cos 2 \beta)}{R \sin 2\beta \; {\rm PP}'} \right],
\end{flalign}
where
\begin{equation}
{\rm PP}'= \sqrt{\dd^2 - 4aR \sin^2 \beta},
\end{equation}
is the distance from point P$(R,\theta,Z)$ to point P$'(a,\theta',z)$. To be rigorous, we should include in the right-hand-side of Eq.(\ref{eq:integkF}) any function of the three variables $\zeta$, $\beta$, and $R$, but this is not necessary here since we work with a definite integral in the following. With $0$ and $a$ as the lower and upper bounds respectively, we get the expression for $\psib(\vec{r})$ defined by Eq.(\ref{eq:psifloat}), and then $\psipod(\vec{r})$ given by Eq.(\ref{eq:psipod}) or Eq.(\ref{eq:psipodbasic}). As $E(\beta,0)=F(\beta,0)=\Pi(\beta,0,0)=\beta$, we deduce that the elliptic integrals cancel each other at the lower bound. We finally get, for $\beta \le \frac{\pi}{2}$:
\begin{flalign}
\label{eq:psifloatfinal}
\psib(\vec{r};a,\beta)  =\dd E(\beta,k) + \frac{a^2-R^2}{\dd}F(\beta,k)  + \frac{a-R}{a+R}\frac{\zeta^2}{\dd} \Pi(\beta,m^2,k)\\ \nonumber
 + R H(\beta) + \zeta T(\beta),
\end{flalign}
where we have defined for convenience:
\begin{equation}
T(\beta) =  - \atan \left[ \frac{\zeta(a+R\cos 2 \beta)}{R \sin 2\beta \; {\rm PP}'} \right] + \atan \left( \frac{\zeta}{\sqrt{\zeta^2+R^2}} \cotan 2\beta \right),
\end{equation}
for the ``trigonometric'' term, and
\begin{flalign}
H(\beta) & =-\sin 2 \beta \left(  \asinh \frac{a+R \cos 2 \beta}{\sqrt{\zeta^2+R^2 \sin^2 2 \beta}} \right. \\ \nonumber
& \qquad \qquad \qquad  \left. - \asinh \frac{R \cos 2 \beta}{\sqrt{\zeta^2+R^2 \sin^2 2 \beta}} \right),
\end{flalign}
for the ``hyperbolic'' term. Note that both $H$ and $T$ depend not only on the amplitude $\beta$ but also on $m$ and $k$ through $a/R$ and $\zeta$. Equation (\ref{eq:psifloatfinal}) is exact. It is valid in all of space, in the plane of the distribution (if $\zeta=0$) as well as off the plane. It can obviously be put in various equivalent forms, in particular by combining the intrinsic functions having two different arguments. As expected, it is an even function of $\zeta$ (the potential is the same above and below the plane of the distribution).

Because of the inverse trigonometric function in Eq.(\ref{eq:daasym}), Eq.(\ref{eq:psifloatfinal}) must not be used for amplitudes $\beta$ larger than $\frac{\pi}{2}$ (unless $T$ must include a constant). Such amplitudes correspond to a floating sector larger than half-a-disc, or $\Delta \theta' > \pi$. This configuration is precisely the one displayed in Fig. \ref{fig:disc3.eps}a. To treat cases with $\beta > \frac{\pi}{2}$, we replace this large floating sector by a circular disc minus a small floating sector. The small floating sector to be removed has the following properties: i) same radius $a$, ii) opening angle $2\pi - \Delta \theta' \le \pi$, and iii) two bounding  angles $\theta''=2\theta-\theta' \ge \theta $ and $\theta'=\theta+\pi \ge \theta''$. The corresponding amplitudes are $\beta''=\frac{\pi}{2} - \frac{\theta-\theta'}{2} \le \frac{\pi}{2}$ and $\beta=0$, respectively. For $\beta > \frac{\pi}{2}$, the potential of the floating sector is then given by:
\begin{equation}
\psib(\vec{r};a,\beta > \frac{\pi}{2})  = \psicd(\vec{r};a) - \psib (\vec{r};a,\beta'' < \frac{\pi}{2}).
\label{eq:psifloatfinalwithbetapiover}
\end{equation}

A typical algorithm to determine $\psib$ whatever $\beta$ by using uniquely Eq.(\ref{eq:psifloatfinal}) is then the following:

\begin{equation}
\boxed{
\begin{array}{l}
{\rm let \,} \alpha \leftarrow \theta'-\theta;\\ 
{\rm let \,} \beta \leftarrow \frac{\pi-\alpha}{2};\\
{\rm if \,} \beta \le \frac{\pi}{2} {\rm \, then} \\
\qquad {\rm let \,}  result \leftarrow \psib(\vec{r};a,\beta);\\
\label{eq:algo}
{\rm  else}\\
\qquad {\rm let \,} \beta' \leftarrow \frac{\pi+\alpha}{2};\\
\qquad {\rm let \,}  result \leftarrow \psib(\vec{r};a,\beta');\\
\qquad {\rm let \,}  result \leftarrow 2 \psib(\vec{r};a,\frac{\pi}{2}) - result;\\
{\rm endif}
\end{array}}
\medskip
\end{equation}

\begin{figure}
\centering
\includegraphics[width=7.5cm,bb=160 90 550 554,clip==, angle=-90]{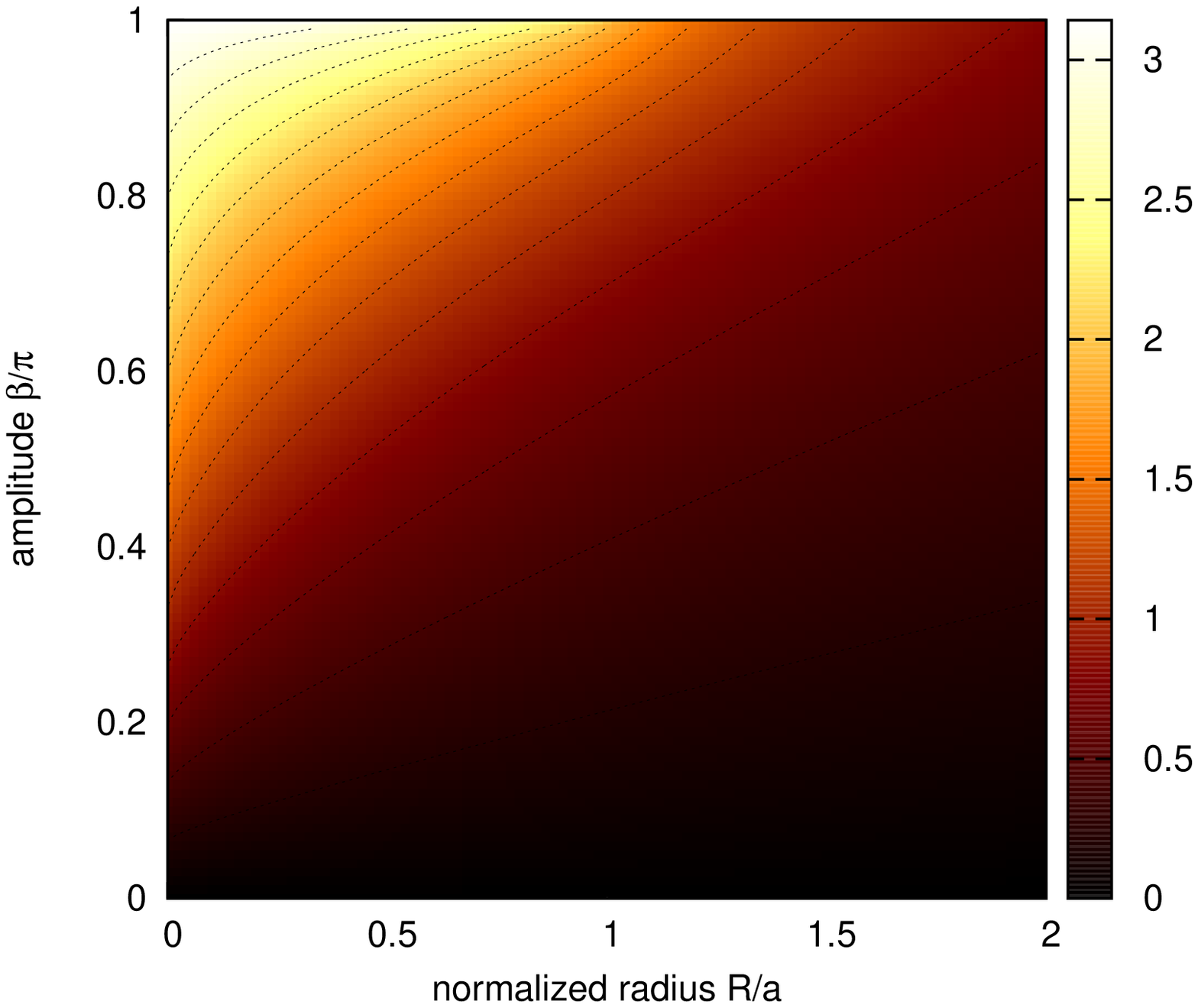}\\\includegraphics[width=7.5cm,bb=160 90 550 554,clip==, angle=-90]{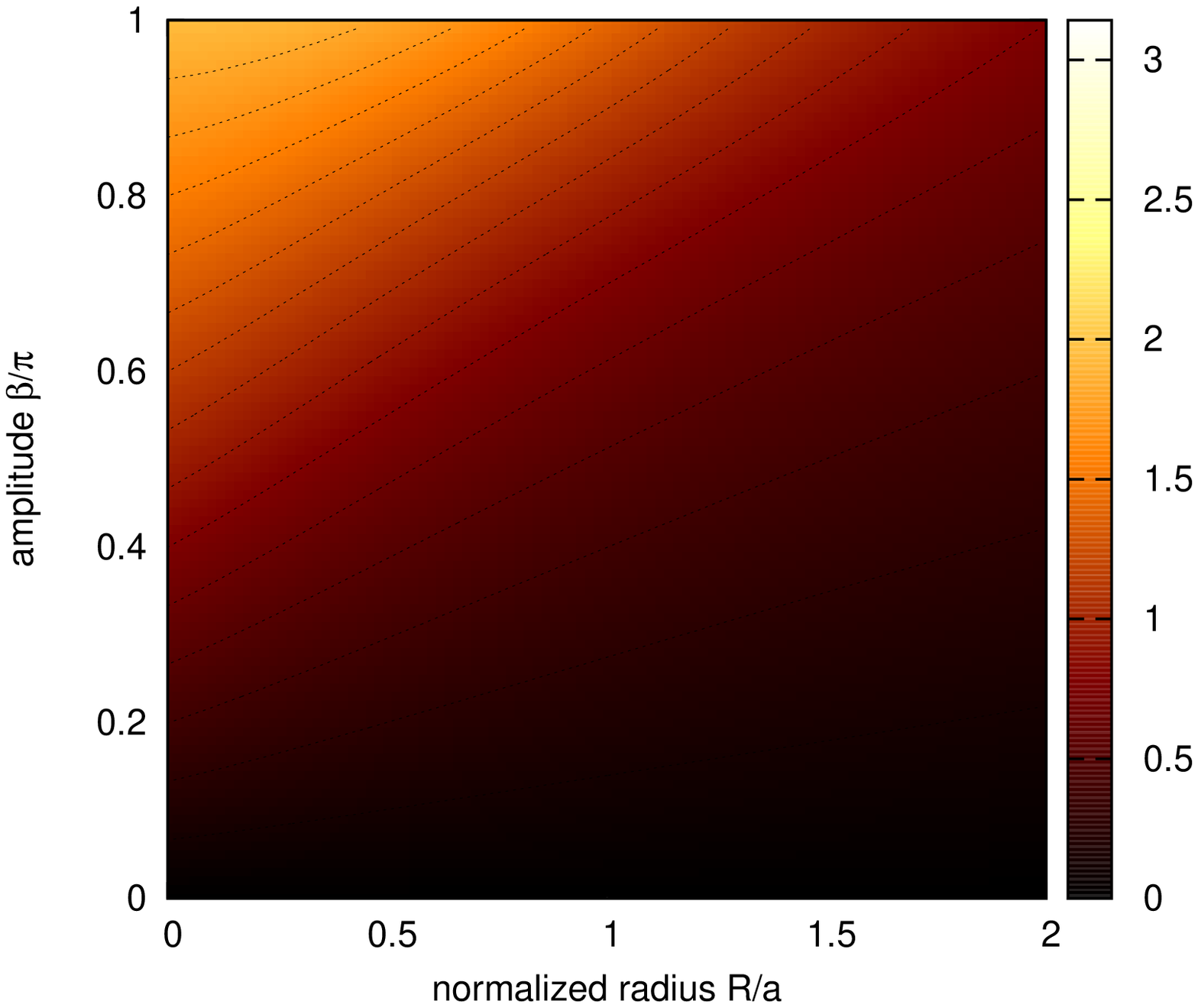}
\caption{Color map showing $\psib(\vec{r},a,\beta)$ from Eq.(\ref{eq:psifloatfinal}) versus $R/a$ and $\beta$ (in units of $\pi$). These cases are for $\zeta=0$ ({\it top panel}) which corresponds to the plane of the floating circular sector, and for $\zeta=a/2$ (off-plane; {\it bottom panel}). The color code is the same. A few contour levels are superimposed}.
\label{fig:plot_beta.ps}
\end{figure}

Figure \ref{fig:plot_beta.ps} displays $\psib(\vec{r};a,\beta$) computed in the $(R/a,\beta)$-plane from Eq.(\ref{eq:psifloatfinal}), for $R/a \in [0,2]$ and $\beta \in [0,\pi]$. When $\beta > \frac{\pi}{2}$, Eq.(\ref{eq:psifloatfinalwithbetapiover}) is used instead. Two cases are considered : $\zeta=0$ (i.e. in the plane of the circular sector), as well as an off-plane case, where $\zeta=a/2$.

\section{Potential of the circular sector and polar cell}
\label{sec:piece}

The potential of a circular sector is found from Eqs.(\ref{eq:psifloatfinal}) and (\ref{eq:psifloatfinalwithbetapiover}) by considering two amplitudes $\beta_1$ and $\beta_2$, according to Eq.(\ref{eq:psipodbasic}). Similarly, the gravitational potential of a polar cell is found from Eq.(\ref{eq:psipodbasic}) by considering two circular sectors having the same bounding angles but two different radii. A typical polar cell is depicted in Fig. \ref{fig:disc3.eps}b. Let $\ain$ be the inner radius of the cell, $\aout$ the outer radius, $\Delta a = \aout-\ain \ge 0$ the radial extension, and $\theta_1'$ and $\theta_2' \ge \theta_1'$ the two bounding angles. So, the potential is:
\begin{equation}
\psicell(\vec{r};\ain,\aout,\beta_1,\beta_2) = \psipod(\vec{r};\aout,\beta_1,\beta_2)-\psipod(\vec{r};\ain,\beta_1,\beta_2),
\label{eq:potpm}
\end{equation}
where, for each circular sector, the quantities $\dd$, $k$ and $m$ must be computed as:
\begin{equation}
\begin{cases}
\dd^2 =(\ain+R)^2+\zeta^2,\\
k=\frac{2\sqrt{\ain R}}{\dd},\\
m=\frac{2\sqrt{\ain R}}{\ain+R}, \qquad   {\rm for } \; a=\ain,
\end{cases}
\end{equation}
and
\begin{equation}
\begin{cases}
\dd^2 =(\aout+R)^2+\zeta^2\\
k=\frac{2\sqrt{\aout R}}{\dd}\\
m=\frac{2\sqrt{\aout R}}{\aout+R}, \qquad  {\rm for } \; a=\aout,
\end{cases}
\end{equation}
while $\zeta$ is the same for the two sectors.

\section{Accelerations}
\label{sec:acceleration}

The gravitational accelerations $-\vec{\nabla} \psipod$ and $-\vec{\nabla} \psicell$ due to the circular sector and polar cell respectively involve the partial derivatives with respect to the three variables $R$, $\theta$ and $Z$. We have
\begin{equation}
\vec{\nabla} \psipod(\vec{r};a,\beta_1,\beta_2) = \vec{\nabla} \psib(\vec{r};a,\beta_1)  - \vec{\nabla}\psib(\vec{r};a,\beta_2),
\end{equation}
and
\begin{equation}
\vec{\nabla} \psicell(\vec{r};\ain,\aout,\beta_1,\beta_2) = \vec{\nabla} \psipod(\vec{r};\aout,\beta_1,\beta_2)-\vec{\nabla} \psipod(\vec{r};\ain,\beta_1,\beta_2),
\end{equation}
where $- \vec{\nabla} \psib(\vec{r};a,\beta)$ is the acceleration of the floating circular sector with bounding angles $0$ and $\theta'= \pi+\theta-2\beta$. The $3$ components of $\vec{\nabla} \psib(\vec{r};a,\beta)$ are derived in Appendices \ref{app:grpod}, \ref{app:gthetapod} and \ref{app:gzpod}. As for the potential (see Eq.(\ref{eq:psifloatfinalwithbetapiover})), the case with $\beta > \frac{\pi}{2}$ must be treated by considering the circular disc, i.e.
\begin{equation}
\vec{\nabla} \psib(\vec{r};a,\beta > \frac{\pi}{2})  = \vec{\nabla} \psicd(\vec{r};a) - \vec{\nabla} \psib (\vec{r};a,\beta'' < \frac{\pi}{2}),
\end{equation}

where $\beta''= \pi - \beta \le \frac{\pi}{2}$, and $\vec{\nabla} \psicd(\vec{r};a) = -\vec{g}^{\rm \, disc}(\vec{r};a)$ (see Appendix \ref{app:g}).

\section{Important remarks and special cases}
\label{sec:remarks}

\paragraph{Case with $\beta = \frac{\pi}{2}$.} As mentioned above, this case corresponds to the value of the potential at the edge of a half-disc (i.e. along the axis defined by $\theta=\theta'$). For that amplitude, all the incomplete elliptic integrals take their complete value. The inverse hyperbolic functions vanish, while the two inverse trigonometric functions may still contribute through an argument eventually infinite. We then need to consider three cases, depending on the position of the observation point relative to the {\it tangent circle}, which is the circle with radius $a$ and tangent to the circular sector. The same kind of distinction exists under axial symmetry \citep{durand64,lassblitzer83}. We have:
\begin{itemize}
\item if $R=a$, then $m=1$ and $a+R\cos 2 \beta= 2 R \cos^2 \beta$. The argument of the first inverse tangent function remains finite. We find:
\begin{equation}
 T(\beta) = - |\zeta|\left[ \atan(\cotan \beta)+\frac{\pi}{2} \right],
\end{equation}
and so
\begin{equation}
\lim_{\beta \rightarrow \frac{\pi}{2}^-} T(\beta) = - |\zeta| \frac{\pi}{2}.
\end{equation}

\item if $R>a$, then $(a-R)/\sin 2 \beta <0$. The two inverse trigonometric functions cancel each other, and so we get:
\begin{equation}
\lim_{\beta \rightarrow \frac{\pi}{2}^-} T(\beta) = 0.
\end{equation}

\item if $R<a$, then $(a-R)/\sin 2 \beta >0$. The two inverse trigonometric functions bring the same contribution and we find:
\begin{equation}
\lim_{\beta \rightarrow \frac{\pi}{2}^-} T(\beta) = - |\zeta| \pi.
\end{equation}

\end{itemize}
We then conclude that:
\begin{equation}
\lim_{\beta \rightarrow \frac{\pi}{2}^-} T(\beta) = - |\zeta| \pi \times
\begin{cases}
1, \quad  {\rm if } \; R<a,\\
\frac{1}{2}, \quad   {\rm if } \; R=a,\\
0, \quad  {\rm if } \; R>a.
\end{cases}
\end{equation}
We see that this result can be written in the compact form:
\begin{equation}
\label{eq:tatpiover2}
\lim_{\beta \rightarrow \frac{\pi}{2}^-} T(\beta) = - |\zeta| \pi \times  \epsilon',
\end{equation}
where $\epsilon'$ is a step-function already defined in Section \ref{sec:disc}. We then fully recover Durand's formula.

\paragraph{Axial symmetry.} Given the above analysis, we see that Durand's formula and Eq.(\ref{eq:psipodbasic}) are indeed equivalent. Actually, axial symmetry implies $\Delta \theta = 2\pi$, which is achieved for instance with $\theta_1'=\theta$ and $\theta_2'=\theta+2\pi$, that is for $2\beta_1 = \pi $ and $2\beta_2 = -\pi$. In particular, the term $- |\zeta| \pi \times  \epsilon'$ in Eq.(\ref{eq:psi_as}) comes from the two trigonometric functions, in the limit $\beta \rightarrow \frac{\pi}{2}^-$. We have:
\begin{flalign}
\psicd(\vec{r};a)& = \psipod(\vec{r};a,\frac{\pi}{2})-\psipod(\vec{r};a,-\frac{\pi}{2})\\\nonumber
& = 2\psib(\vec{r};a,\frac{\pi}{2}),
\end{flalign}
which is nothing but the potential of {\it two half-discs}, hence the factor $2$ in Durand's formula.

\paragraph{Calculations in the plane.} For $\zeta=0$, we have $k=m$ and $\dd=(a+R)$. The two functions $\Pi$ and $T$ do not contribute. Equation (\ref{eq:psifloat}) then becomes:
\begin{flalign}
\label{eq:psib_plane}
\psib(\vec{r}=R\vec{e}_R;a,\beta)&  =(a+R)E(\beta,m) + (a-R)F(\beta,m)\\ \nonumber
& -  R \sin 2\beta \left(  \asinh \frac{a+R \cos 2 \beta}{R|\sin 2 \beta|}  -\asinh \frac{\cos 2 \beta}{|\sin 2 \beta|} \right).
\end{flalign}

\paragraph{Polar axis.} We see from  Eq.(\ref{eq:psib_plane}) that for $R=0$ and $\zeta=0$, $\dd=a$, and $k=m=0$. Only the incomplete integral $E$ and $F$ contribute, and they are identical to the amplitude. The potential of the floating sector (up to a factor $-G\Sigma$) is then given by:
\begin{equation}
\psib(\vec{0};a,\beta)=2a\beta,
\end{equation}
which is nothing but the length of the arc. So, the potential of the circular sector {\it at its corner} is:
\begin{equation}
\psipod(\vec{0};a,\beta_1,\beta_2)=2a(\beta_1-\beta_2) = a \Delta \theta'.
\end{equation}
For $\zeta \ne 0$, we have :
\begin{equation}
T(\beta) = - 2\beta |\zeta|,
\end{equation}
and then
\begin{equation}
\psib(\vec{r}=Z\vec{e}_Z;a,\beta) = 2 \beta \left(\sqrt{a^2+\zeta^2} - |\zeta| \right),
\end{equation}
which leads to:
\begin{equation}
\psipod(\vec{r}=Z\vec{e}_Z;a,\beta_1,\beta_2)= \left(\sqrt{a^2+\zeta^2} - |\zeta| \right)\Delta \theta'.
\end{equation}

\paragraph{Long-range behavior.} At large distance from the system, the modulus $k \rightarrow 0$, meaning that $F(\beta,k) \approx \beta$ to second-order in $k$. At the same order, we then have
\begin{equation}
\psib(\vec{r};a,\beta) \approx \int_0^a{\frac{2a}{\delta}\beta da}.
\end{equation}
With $\delta \approx r \gg \sqrt{a^2+z^2}$, $r$ being the spherical radius, we have
\begin{equation}
\psib(\vec{r};a,\beta) \approx \frac{1}{r}\int_0^a{\frac{2a} \beta da} = \frac{1}{r} \beta a^2.
\end{equation}
By difference, we get from Eq.(\ref{eq:psipodbasic}):
\begin{flalign}
\psipod(\vec{r};a,\beta_1,\beta_2) &\approx \frac{a^2}{r}(\beta_1-\beta_2)= \frac{a^2}{2r}\Delta \theta'= \frac{M}{r},
\end{flalign}
which is the expected result: far enough from the system, the potential is proportional to the mass and inversely proportional to the relative distance.

\begin{figure}
\centering
\includegraphics[width=7.5cm,bb=160 90 550 554,clip==, angle=-90]{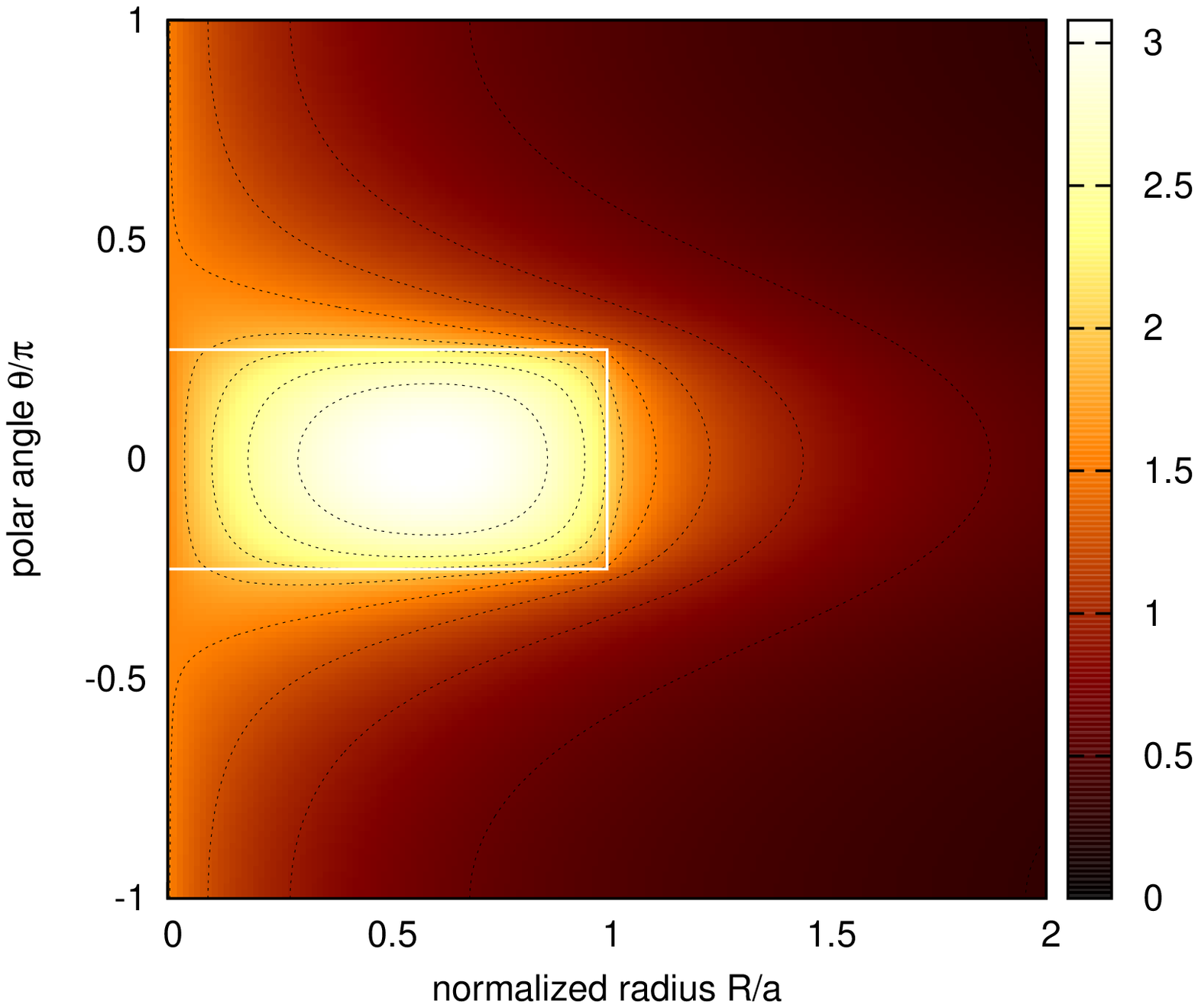}\\
\includegraphics[width=7.5cm,bb=160 90 550 554,clip==, angle=-90]{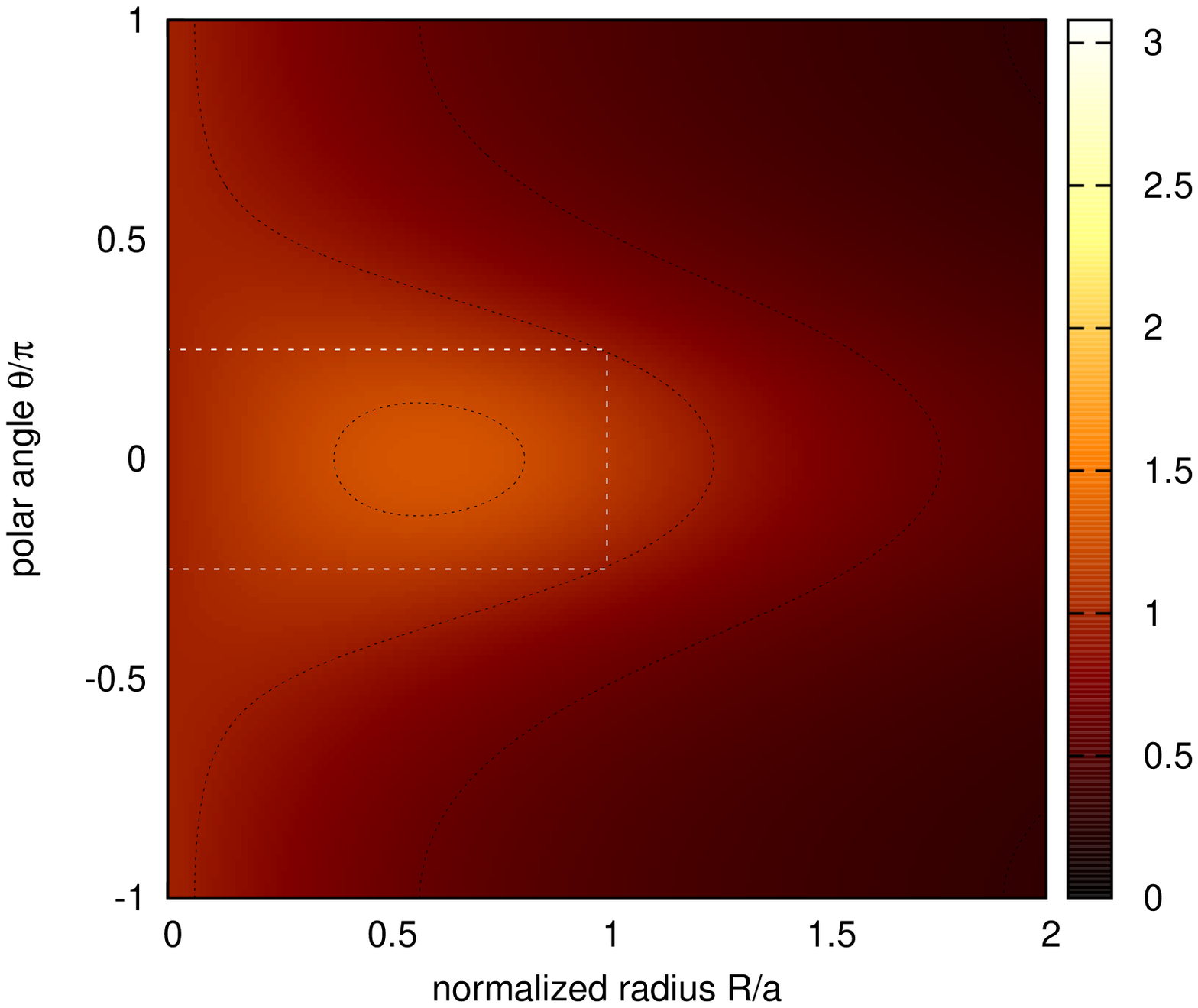}
\caption{Color map showing the potential $\psipod$ of the circular sector (the boundary is shown in white) with opening angle $\Delta \theta' = \frac{\pi}{2}$: in the plane ({\it top panel}) and off-plane with $\zeta=a/2$ ({\it bottom panel}). The color code is the same. A few contour levels are superimposed.}
\label{fig:plot_pod.ps}
\end{figure}
\begin{figure}
\centering
\includegraphics[width=7.5cm,bb=160 90 550 554,clip==, angle=-90]{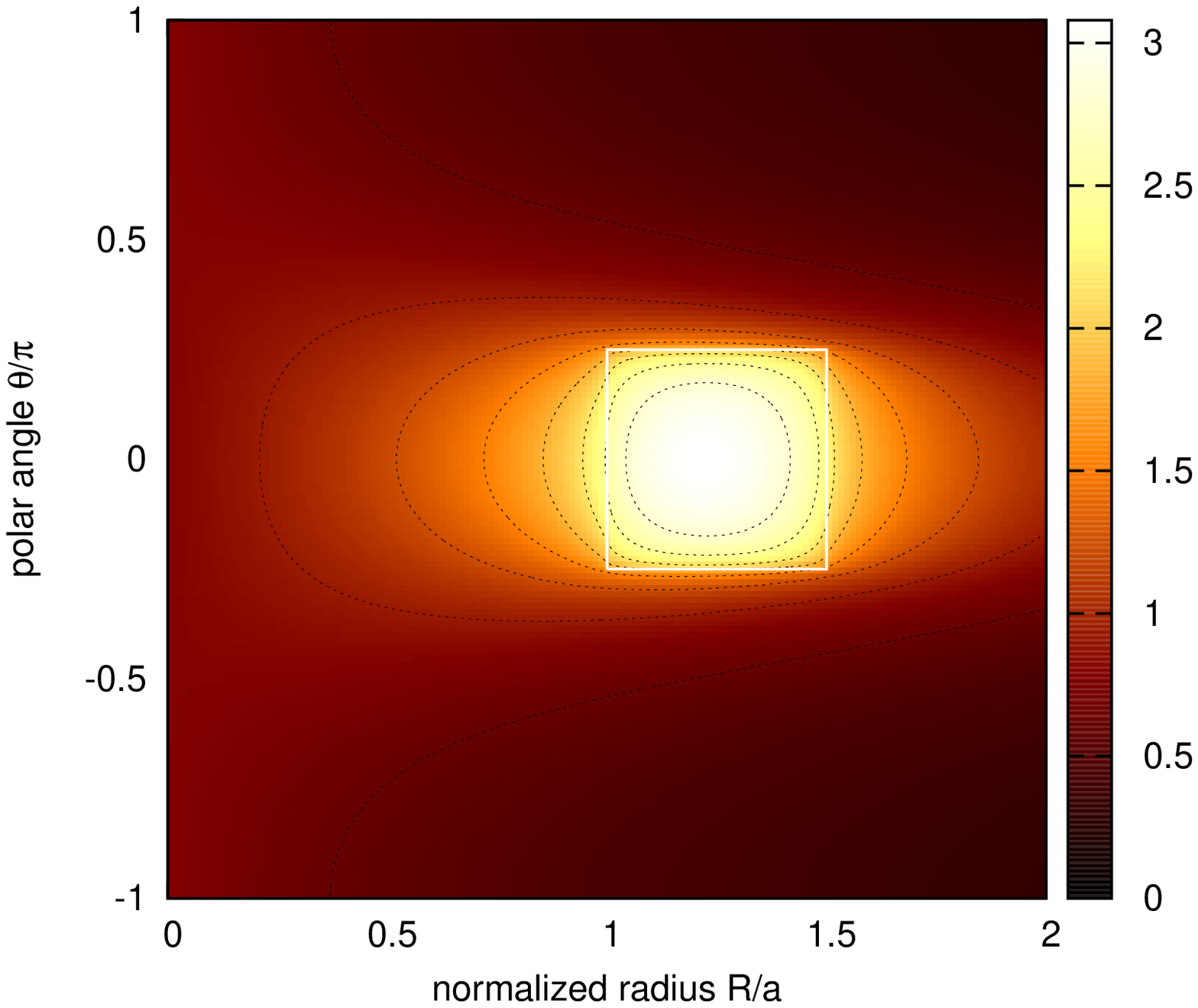}\\
\includegraphics[width=7.5cm,bb=160 90 550 554,clip==, angle=-90]{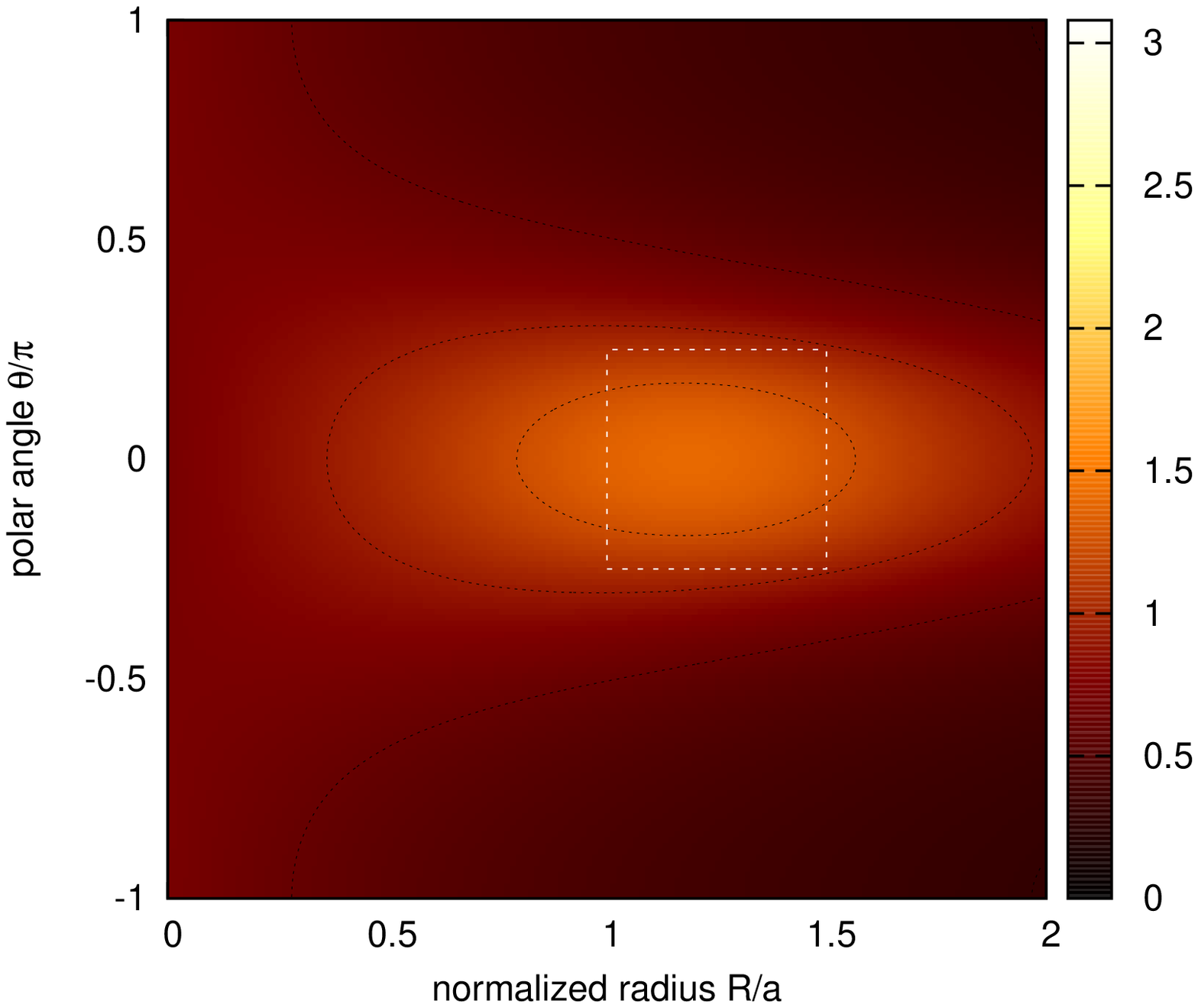}
\caption{Same conditions as for Fig. \ref{fig:plot_pod.ps} but for a polar cell with outer radius $\aout=\frac{3}{2}\ain$, inner radius $\ain=1$ and opening angle $\Delta \theta' = \frac{\pi}{2}$ (the boundary is shown in white).}
\label{fig:plot_pcell.ps}
\end{figure}

\section{A few numerical examples}
\label{sec:examples}

The way to compute the potential depends mainly on the polar angle $\theta$ relative to $\theta_1'$ and $\theta_2'$. We recommend to proceed as follows: 
\begin{itemize}
\item if $\theta_2'>\theta_1' \ge \theta$,  the two amplitudes $\beta_1$ and $\beta_2$ are less than $ \frac{\pi}{2}$. The potential is then computed directly from Eq.(\ref{eq:psipodbasic}) where each term is determined from Eq.(\ref{eq:psifloatfinal});
\item if $\theta_1'<\theta_2' \le \theta$, we can proceed as above by considering a new observation point P''($\vec{r}''$) mirror symmetric of P with respect to the plane of the distribution for which the potential is the same. The altitude of this point is $2z-Z$ (the radius and polar angle are unchanged). The new bounding angles are : $\theta_1''=-\theta_2'$ and $\theta_2''=-\theta_1'$. The potential is then
\begin{equation}
\psipod(\vec{r};\beta_1,\beta_2)=\psipod(\vec{r}'';\beta_1',\beta_2'),
\end{equation}
where $2\beta' = \pi -(\theta''-\theta)$.
\item if $\theta_2' \ge \theta > \theta_1$, then $\beta_2 \le \frac{\pi}{2} \le \beta_1$. In this case, $\psib (\vec{r};a,\beta_2)$ is determined from Eq.(\ref{eq:psifloatfinal}) and $\psib (\vec{r};a,\beta_1)$ is found from Eq.(\ref{eq:psifloatfinalwithbetapiover}). We finally have (see also Eq.(\ref{eq:algo})):
\begin{equation}
\psipod(\vec{r};a,\beta_1,\beta_2)  =  \psicd(\vec{r};a) - \psib(\vec{r};a,\beta_1') -  \psib(\vec{r};a,\beta_2).
\end{equation}
where $\beta_1'=\pi-\beta_1$.
\end{itemize}

Figure \ref{fig:plot_pod.ps} shows the potential of a circular sector in two $(R,\theta)$-planes. The first panel is for $ \zeta=0$ (i.e. in the plane of the cell), and the second panel is for $\zeta=a/2$. The opening angle of the sector is $\Delta \theta' = \frac{\pi}{2}$, i.e. $\theta_1'=-\theta_2'=-\frac{\pi}{4}$. Finally, the potential of a polar cell computed from Eq.(\ref{eq:potpm}) is shown in Fig. \ref{fig:plot_pcell.ps} for $\zeta=0$ and for $\zeta=a/2$ (off-plane). The opening angle is the same, the radius at the outer edge is $\aout=\frac{3}{2}\ain$ and $\ain=1$. 

\begin{figure}
\centering
\includegraphics[width=8.5cm,bb=16 11 736 590,clip==]{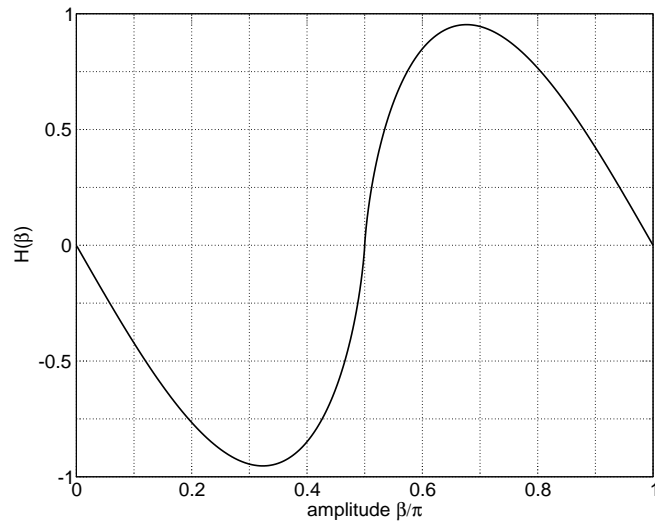}
\caption{The odd function $H(\beta)$ for $a=R$ and $\zeta=0$ versus $\beta$ (in units of $\pi$).}
\label{fig:h.eps}
\end{figure}

\section{A theorem}
\label{sec:theorem}

In the plane of the distribution, Eq.(\ref{eq:psifloatfinal}) takes a particularly simple form since $m=k$ (i.e. $\zeta=0$). In addition, for $R=a$ we have $m=k=1$. All along this tangent circle, we have:
\begin{equation}
\label{eq:psipodfinalata}
\psib(\vec{r}=a \vec{e}_R;a,\beta)  = 2a E(\beta,1) + R H(\beta),
\end{equation}
where
\begin{flalign}
H(\beta)&=-\sin 2\beta \left( \asinh \frac{1+ \cos 2 \beta}{| \sin 2 \beta |} - \asinh \frac{\cos 2 \beta}{| \sin 2 \beta |} \right),\\\nonumber
&= - \sin 2\beta \times \ln \left(1+\frac{1}{|\cos \beta|} \right)
\end{flalign}
The function $H(\beta)$ is plotted versus $\beta$ in Fig. \ref{fig:h.eps} for $a=R$ and $\zeta=0$. It is an odd function of the angle $2\beta$, and is $2\pi$-periodic in this same angle. As a consequence, when integrated over any interval $[0,\pi]$, we always get:
\begin{equation}
\int_0^{\pi}{H(\beta)d\beta}=0.
\label{eq:h11}
\end{equation}
In fact, this periodicity does not depend on $a$ and $R$, and so Eq.(\ref{eq:h11}) is true even for $m \ne 1$. We can easily integrate over one period each term in the right-hand-side of Eq.(\ref{eq:psipodbasic}). We take the interval $[\theta_1'-\pi,\theta_1'+\pi]$, which means $2 \beta_1 \in [0,\pi]$. As we have
\begin{equation}
E(\beta,1) =
\begin{cases}
\sin \beta \quad   {\rm if } \beta \in [-\frac{\pi}{2},\frac{\pi}{2}],\\\\
2-\sin \beta \quad   {\rm if }  \beta \in [\frac{\pi}{2},\frac{3\pi}{2}],
\end{cases}
\end{equation}
we find:
\begin{flalign}
\int_{\theta_1'-\pi}^{\theta_1'+\pi}{\psib(\vec{r} = a \vec{e}_R;a,\beta_1) d\theta} & = 2 \int_0^{\pi}{\psib(\vec{r} = a \vec{e}_R;a,\beta_1) d\beta_1}\\ \nonumber 
& = 4 \pi a.
\end{flalign}
Regarding the second term, the integration which is to be performed over {\it the same interval} in $\theta$, gives:
\begin{equation}
\int_{\theta_1'-\pi}^{\theta_1'+\pi}{\psib(\vec{r}=a \vec{e}_R;a,\beta_2) d\theta} = 4a ( \pi - \Delta \theta').
\end{equation}
Consequently, the potential of the circular sector integrated over the whole tangent circle is
\begin{equation}
\int_{2\pi}{\psipod(\vec{r}=a \vec{e}_R;a,\beta_1,\beta_2) d\theta} = 4a \Delta \theta'.
\label{eq:avpot}
\end{equation}
We can therefore postulate that {\it the angular average  of the potential along the tangent circle of any circular sector with radius $a$ and opening angle $\Delta \theta'$ is} (up to a factor $-G \Sigma$):
\begin{equation}
\frac{1}{2\pi}\int_{2\pi}{\psipod(\theta) d\theta} = 4a f ,
\end{equation}
where
\begin{equation}
f=\frac{\Delta \theta'}{2\pi} \in [0,1]
\end{equation}
is the {\it opening factor} of the circular sector. This relation can be rewritten into different forms. In particular, by using the value at $R=0$ (see Sect. \ref{sec:remarks}), we also have:
\begin{equation}
\frac{1}{2\pi}\int_{2\pi}{\psipod(\theta) d\theta} = \frac{2}{\pi}\psipod(\vec{0}).
\end{equation}
meaning that the {\it angular average of this potential is $\frac{2}{\pi}$ times the value at the corner}. For the circular disc, we have $\Delta \theta'=2\pi$ and so
\begin{equation}
\frac{1}{2\pi}\int_{2\pi}{\psicd(\theta) d\theta} = 4a.
\end{equation}

\begin{figure}
\centering
\includegraphics[width=8.5cm,bb=16 11 740 583,clip==]{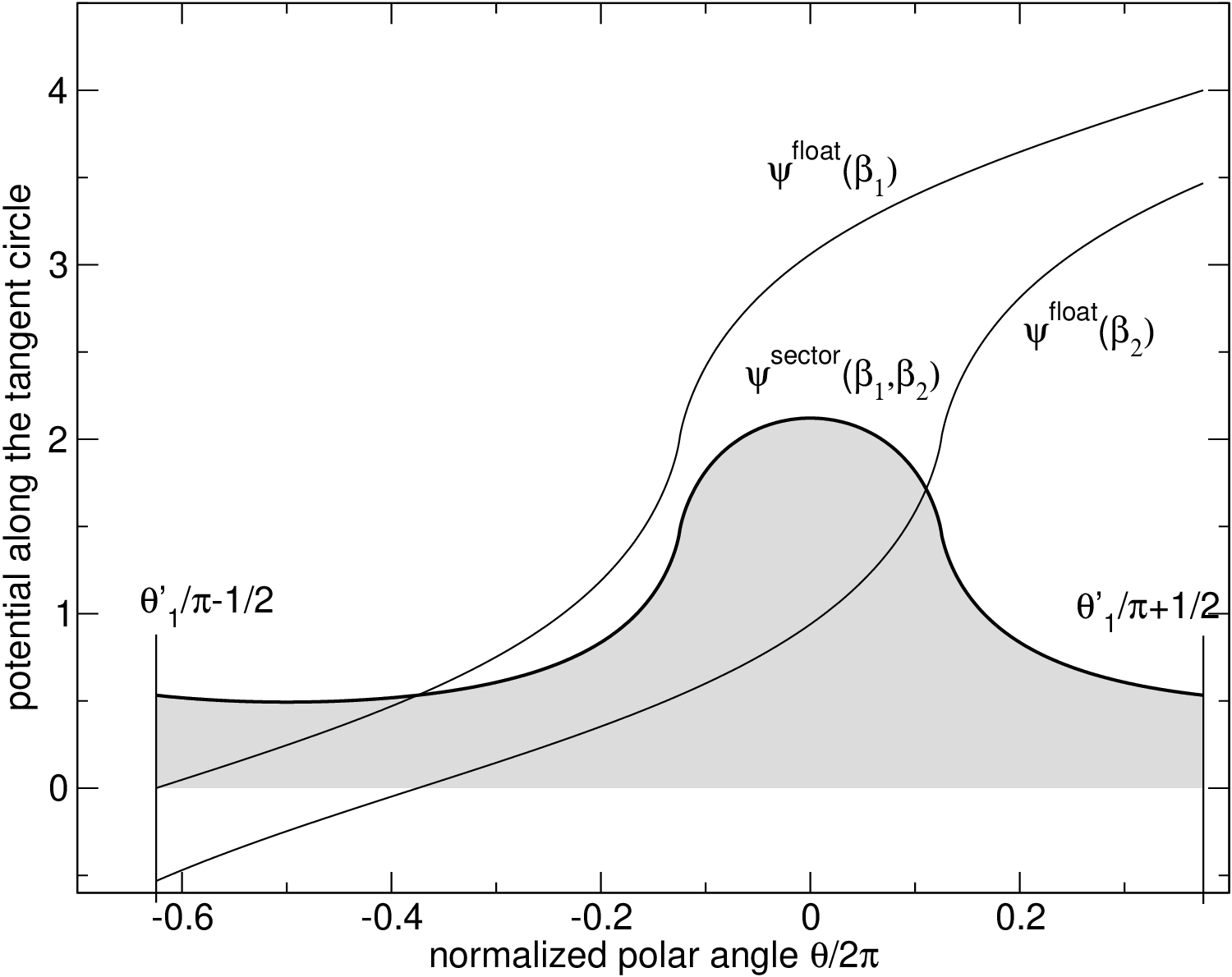}\\
\includegraphics[width=8.5cm,bb=16 11 740 583,clip==]{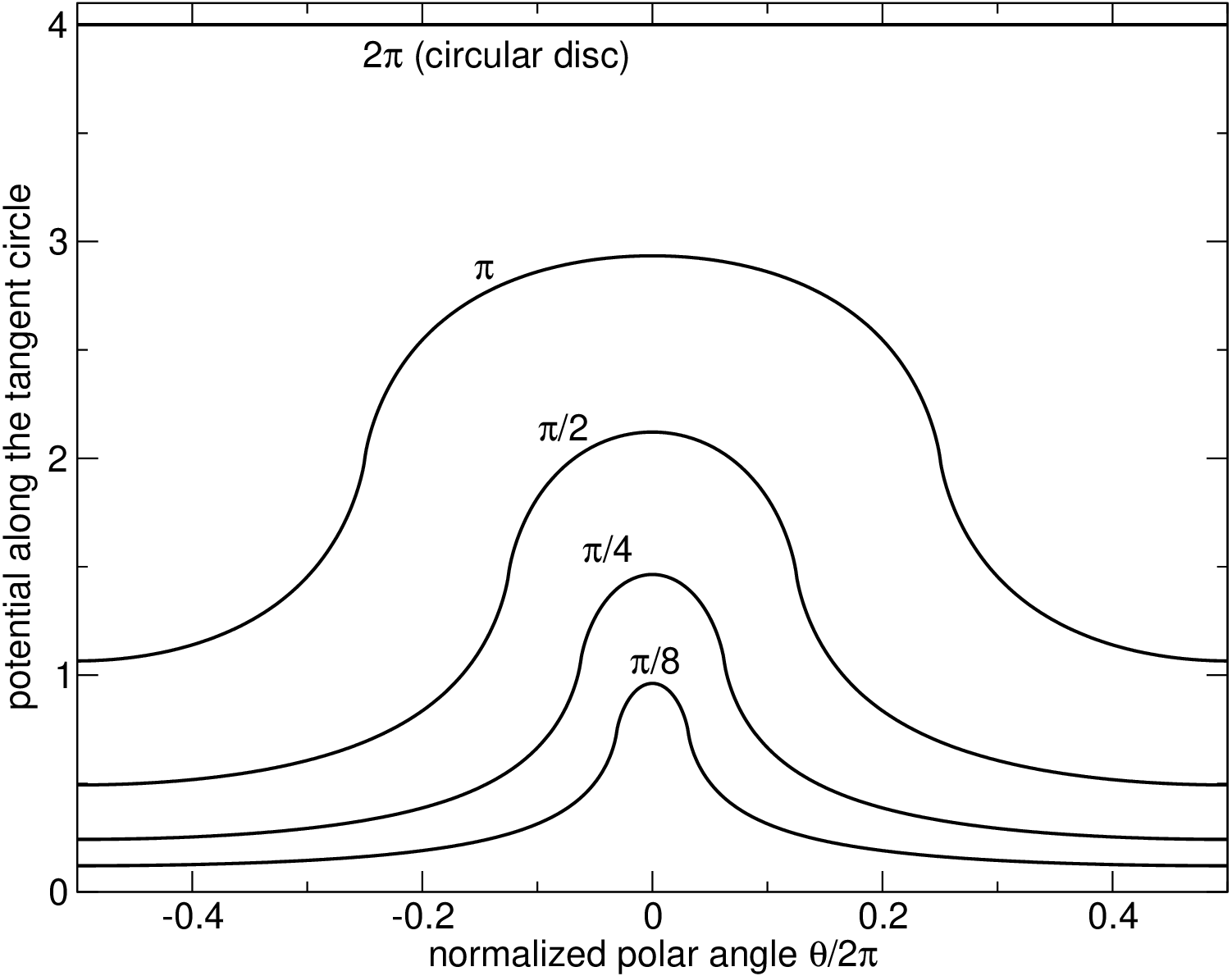}
\caption{The potentials $\psib(\vec{r}=a \vec{e}_R;a,\beta_1)$ and  $\psib(\vec{r}=a \vec{e}_R;a,\beta_2)$ as well as their difference $\psipod$ versus $\theta$ along the tangent circle ({\it top panel}). Here, the parameters are $a=1$ and $\theta_2'=\frac{\pi}{4}=-\theta_1'$, and so $\Delta \theta'=\frac{\pi}{2}$ and $f=\frac{1}{4}$. According to Eq.(\ref{eq:avpot}), the area (shown in grey) under this last curve is $a$. The potential $\psipod$ versus the polar angle $\theta$ along the circle tangent to the circular sector, for different opening angles $\Delta \theta'$ labelled on the curves, and $a=1$ ({\it bottom panel}). The area under each curve is $4a \Delta \theta'$ according to Eq.(\ref{eq:avpot}).}
\label{fig:psiata.eps}
\end{figure}

Figure \ref{fig:psiata.eps}a shows $\psib$ versus $\theta$, as well as the potential of the corresponding circular sector $\psipod$. Figure \ref{fig:psiata.eps}b shows $\psipod$ versus $\theta$ over one period for various opening angles $\Delta \theta'$.

It is probably possible to derive other theorems by considering only the range $[\theta_1',\theta_2']$ (instead of a full period), or integration along the radius.

\section{Curvature effects and high resolutions}
\label{sec:curvature}

When the opening angle $\Delta \theta'$ of the polar cell tends to zero, curvature effects are expected to gradually vanish. The cell becomes more and more rectangular, i.e. Cartesian. It is interesting to analyze how potential values vary in such a case. In particular, it can be interesting to be able to decide whether or not the formula for the Cartesian (i.e. rectangular), homogeneous cell can be used instead of the formula for the polar cell, for a given accuracy, and to quantify the deviation between the two. This is what we have done by considering practical cases. In current numerical simulations of self-gravitating discs, the discretisation consists in a grid with $N \times L$ cells ($N$ is for the radial direction, and $L$ is for the angular direction), typically a few hundred in each direction \citep[e.g.][]{nelson06,bm08,pierens11}. The angular resolution $\frac{1}{L}$ is often smaller than the radial resolution $\frac{1}{N}$, in order to produce, at least at some radius, polar cells with almost square shape\footnote{This preserves a certain isotropy of the precision of the differentiation schemes of fluid equations.}. This is achieved when:
\begin{equation}
a \Delta \theta' \sim \Delta a,
\end{equation}
where $\Delta \theta'$ is the opening angle (see Sec. \ref{sec:strategy}), and $\Delta a$ is the radial extension (see Sec \ref{sec:piece}). Since, by nature, cells have not the same size in polar coordinates, while the shapes stay homothetic, this conditions cannot be met everywhere. For a regular radial discretisation, this condition translates into:
\begin{equation}
a \frac{2\pi}{L} \sim \frac{\aout-\ain}{N},
\end{equation}
and we see that the polar cell is close to the square cell $L \approx 6N$, while at twice this distance, this is achieved for  $L=3N$.

\begin{figure}
\centering
\includegraphics[width=7.5cm,bb=160 90 550 554,clip==,angle=-90]{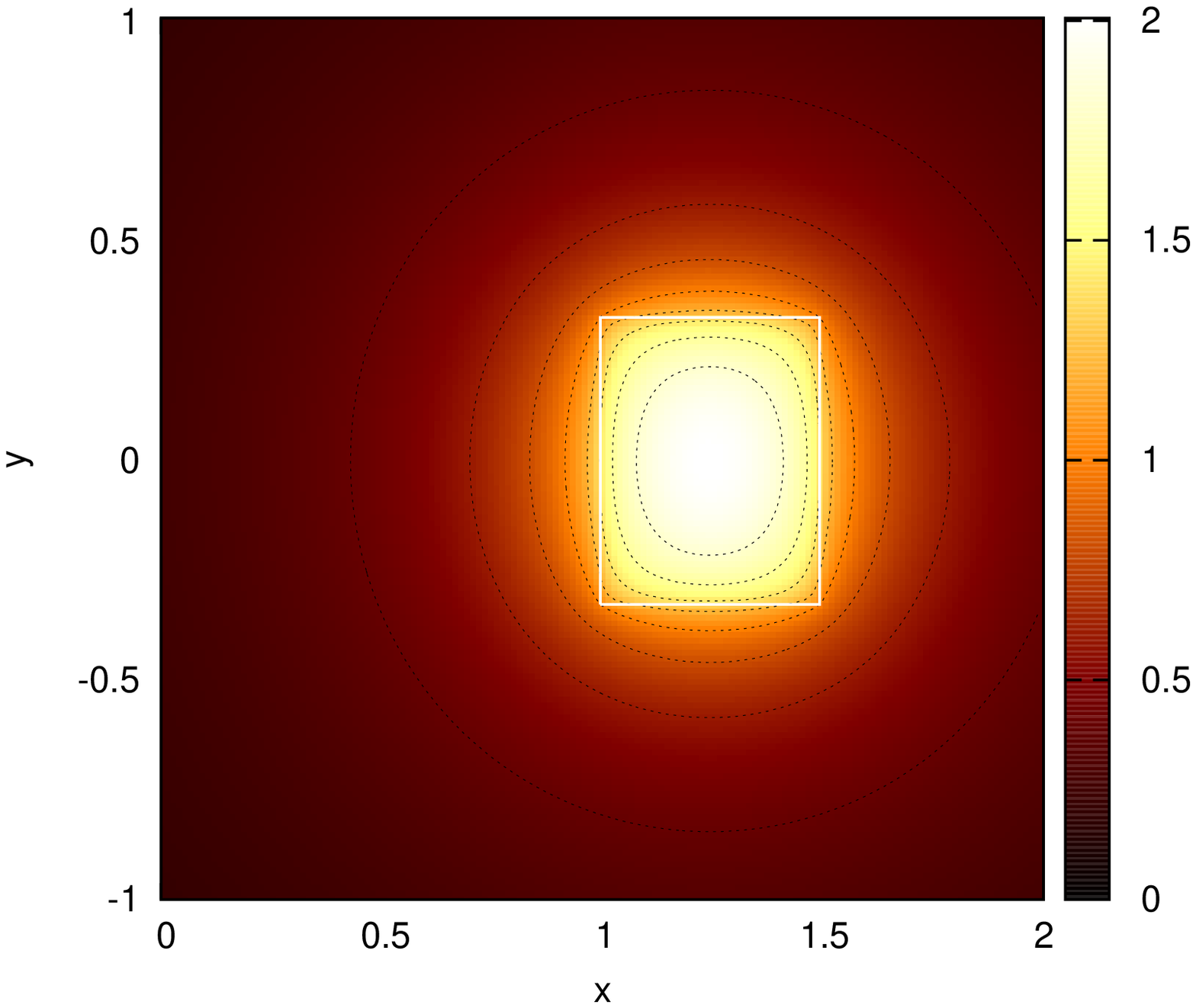}\\
\includegraphics[width=7.5cm,bb=160 90 550 554,clip==,angle=-90]{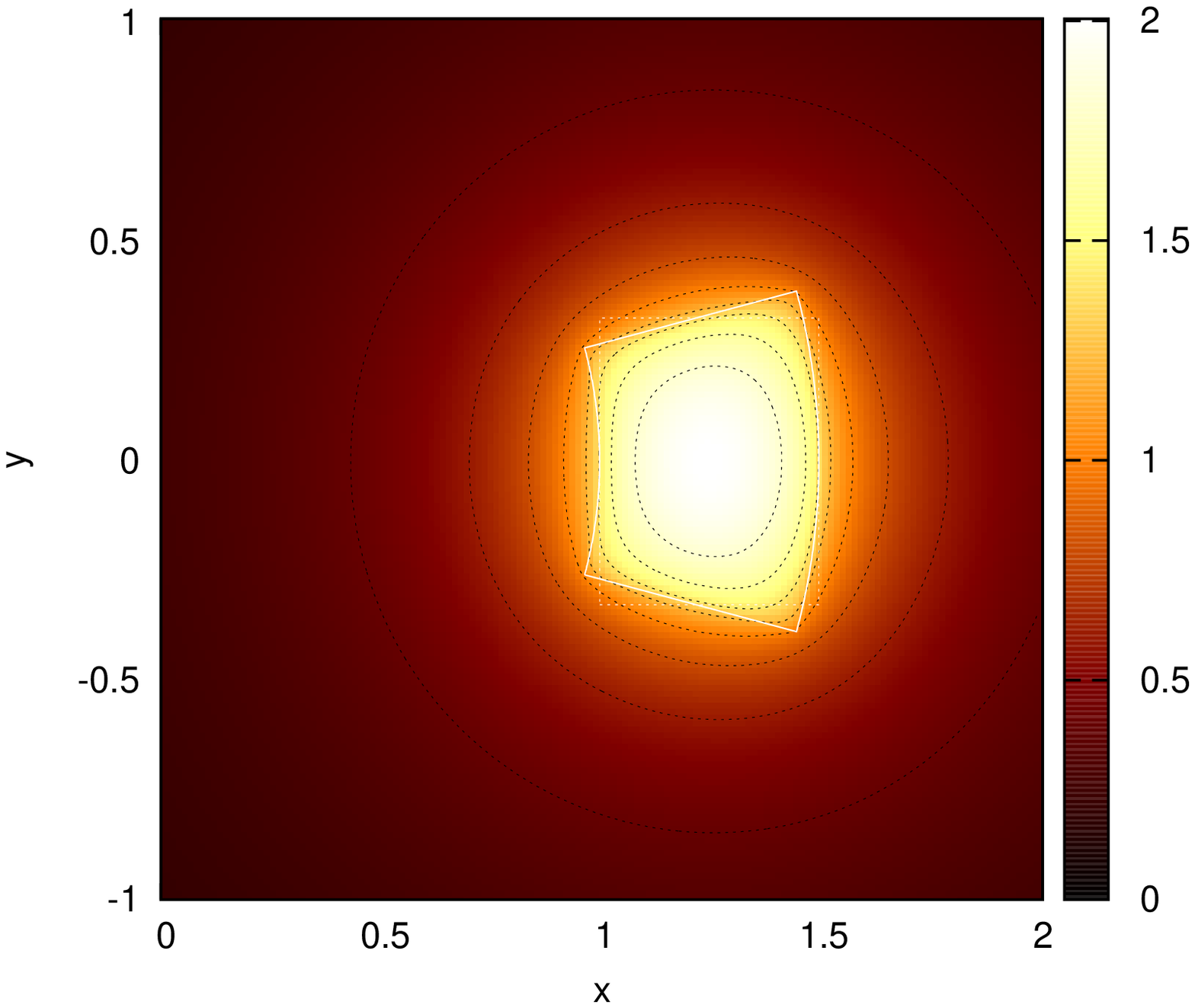}
\caption{Color map showing the potential $\psirect$ of a rectangular sheet in the $(x,y)$-plane computed from Eq.(\ref{eq:psi_rectplate}) ({\it top panel}). The four corners have Cartesian coordinates $(1,y_1')$, $(\frac{3}{2},y_1')$, $(1,-y_1')$ and $(\frac{3}{2},-y_1')$, where $y_1'$ is such that the mass of the rectangular cell is the same as the mass of the polar cell, according to Eq.(\ref{eq:analog}c). The boundary of the Cartesian cell is shown as a white line. A few contour levels are superimposed. Same legend but for the potential $\psicell$ of a polar cell computed from Eq.(\ref{eq:potpm}) ({\it bottom panel}). The parameters are $\zeta=0$, $\ain=1$, $\aout=\frac{3}{2}$ and $\Delta \theta'=\pi/6$, i.e. $N=2$ and $L=6$ according to Eq.(\ref{eq:setup}). The boundary of the Cartesian cell with same mass is recalled (dotted line).}
\label{fig:psirect.ps}
\end{figure}

\begin{figure}
\centering
\includegraphics[width=8.5cm,bb=150 50 580 626,clip==,angle=-90]{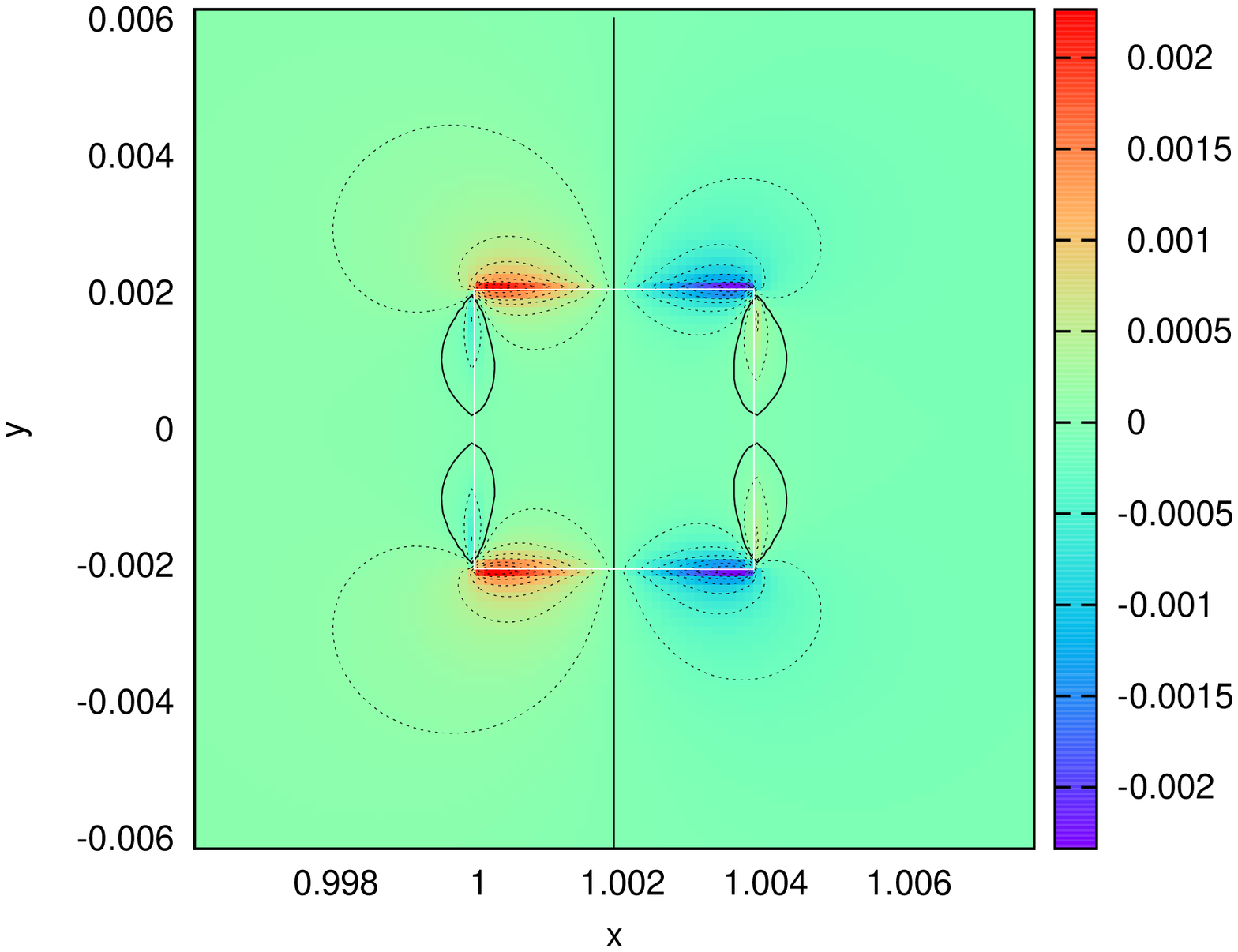}
\caption{Color map showing the relative deviation $\Delta$ defined by Eq.(\ref{eq:deltapsirect}) between the potential of the polar cell and the potential of the Cartesian cell with same mass. The parameters are $\ain=1$, $N=256$ and $L=6N$ (see also Figs. \ref{fig:psirect.ps}a and \ref{fig:psirect.ps}b). A few contour levels are superimposed.}
\label{fig:psidiff.ps}
\end{figure}

Like a polar cell, a Cartesian cell is defined by four corners with coordinates $(x_1',y_1')$, $(x_2',y_1')$, $(x_2',y_2')$ and $(x_1',y_2')$ and its mass is $M = (x_2'-x_1')(y_2'-y_1')$, per unit surface density. The expression for the potential of such a sheet, denoted $\psirect$, is known \citep[e.g.][]{durand64}; it is recalled in the Appendix \ref{app:rectplate}. The potential of a rectangular sheet with $x_2'-x_1'=y_2'-y_1'=\frac{1}{2}$ is shown in Fig. \ref{fig:psirect.ps}a. We have then compared the potential of the polar cell $\psicell$ and the potential of the Cartesian cell $\psirect$ in the case where the two systems have the same mass, for various numerical resolutions $1/N$ and $1/L$. To complete the above example, we show in Fig. \ref{fig:psirect.ps}b the potential of the polar cell having the same mass and same radial extension as the rectangular sheet. We have compared the two potentials numerically. For this purpose, the setup is the following:
\begin{equation}
\begin{cases}
\ain=1,\\
\aout=1+\frac{1}{N}\\
\theta_1'=-\frac{2\pi}{2L}\\
\theta_2'=-\theta_1'
\end{cases}
\label{eq:setup}
\end{equation}
for the polar cell, and
\begin{equation}
\begin{cases}
x_1'=\ain,\\
x_2=\aout\\
y_1'=-\frac{M}{2(x_2'-x_1')}\\
y_2'=-y_1'
\end{cases}
\label{eq:analog}
\end{equation}

for the rectangular sheet. The value of $y_1'$ ensures that the mass of the two systems remains the same whatever the parameters $N$ and $L$. The relative difference
\begin{equation}
\Delta \equiv 1-\frac{\psicell}{\psirect}
\label{eq:deltapsirect}
\end{equation}
is shown in Fig. \ref{fig:psidiff.ps} for $N=256$ and $L=6N$ inside the cell and close neighbourhood. As expected, the largest deviations are observed {\it at the four corners and along the boundaries}, in excess or in default. The maximum value is $\Delta_{\rm max} \approx 2 \times 10^{-3}$ here (in absolute). The deviations remain globally small : i) along the mean arc $R=\frac{1}{2}(\ain+\aout)$, and ii) along the symmetry axis $\theta=\frac{1}{2}(\theta_1'+\theta_2')$. This means that the {\it radial and angular accelerations and forces for the rectangular sheet and for the polar cell} are close to each other there. The deviation at the center of the cell $\Delta_{\rm c}$ is particularly small, with  $\Delta_{\rm c} \approx 10^{-7}$.

\begin{figure}
\centering
\includegraphics[width=9.5cm,bb=17 46 767 595,clip==]{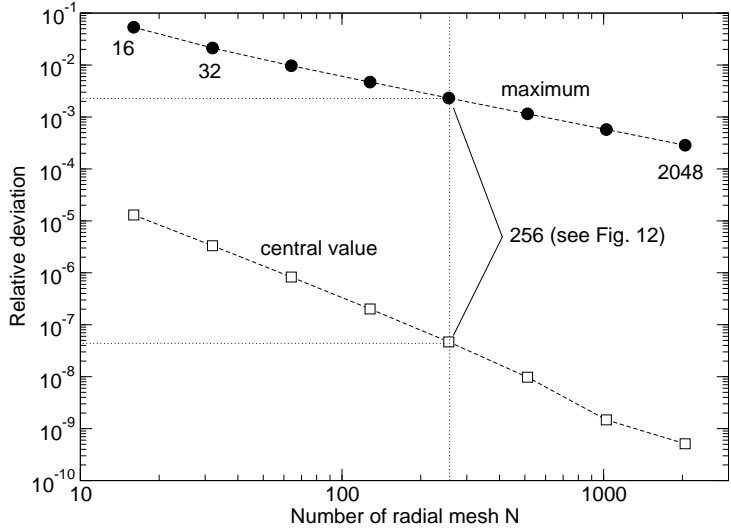}
\caption{Relative deviation $\Delta$ between $\psicell$ and $\psirect$ as a function of the resolution parameters $N$ and $L=6N$: at the center of the polar cell ({\it white square}) and the maximum value ({\it filled circles}) taken over the whole potential map (see also Fig.\ref{fig:psidiff.ps} for $N=256$).}
\label{fig:maxratio.eps}
\end{figure}

Figure \ref{fig:maxratio.eps} shows $\Delta_{\rm max}$  and $\Delta_{\rm c}$ for $N$ running from $2^4=16$ to $2^{12}=4096$ and $L=6N$ still. Both $\log | \Delta_{\rm max} |$ and $\log |\Delta_{\rm c} |$ vary as a power-law of $N$. There is about $4$ orders of magnitude between the deviation at the corners and the deviation at the center. We have roughly:
\begin{equation}
| \Delta_{\rm max} | \approx 2 \times 10^{-3} \times \left(\frac{256}{N}\right)^{1.1}
\end{equation}
for boundary values, and
\begin{equation}
| \Delta_{\rm c} | \approx 4 \times 10^{-8} \times \left(\frac{256}{N}\right)^{2.1}
\end{equation}
for values at the cell center. We have considered different ratios $L/N$, in particular $L=3N$ and $L=12N$ which corresponds to polar cells slightly elongated in the angular and in the radial directions, respectively. We have noted that all the results are globally similar.

\section{Approximation in the vicinity of the cell}
\label{sec:ax}

As a matter of fact, we can find a good approximation for the potential in the vicinity of the polar cell. This may actually be interesting not to manipulate incomplete elliptic integrals and to work with a less complicated expresssion. The vicinity of the polar cell means $\beta$-amplitudes close to $\frac{\pi}{2}$ and moduli $m$ close to unity. In such a case, we can use the following expansion based on the work by \cite{vdv69}:
\begin{equation}
\begin{cases}
E(\beta,m) \approx 1,\\
F(\beta,m) \approx \ln \frac{4 \tan \beta}{1+\sqrt{1+{m'}^2 \tan^2 \beta}} =  \ln \frac{4 (2+\varpi) \sin \beta}{(2+\varpi) \cos \beta + W}
\end{cases}
\label{eq:axef}
\end{equation}
where $m'=\sqrt{1-m^2}$ is the parameter complementary to $m$,
\begin{equation}
W(\varpi,\beta)=\sqrt{4(1+\varpi)\cos^2 \beta +\varpi^2},
\end{equation}
and
\begin{equation}
\varpi=\frac{R-a}{a}.
\end{equation}
In contrast, it does not seem possible to approximate the hyperbolic term $H(\beta)$ for $m=k$ without generating noticeable errors. So we take its exact form:
\begin{equation}
%H(\beta) = -\sin 2\beta \times \ln \left(  1+ \frac{2}{\varpi + \sqrt{4(1+\varpi)\cos^2 \beta +\varpi^2}}\right)
H(\beta) = -\sin 2\beta \times \ln \left[ 1+ \frac{2}{W(\varpi,\beta)}\right]
\end{equation}

%We see that $W(1,\beta)=| \cos \beta |$, in agreement with the expression established in Sect. \ref{sec:theorem}.
The approximation for the potential of the floating sector, denoted $\psib_0$, is found from Eq.(\ref{eq:psifloatfinal}) or Eq.(\ref{eq:psib_plane}) using Eq.(\ref{eq:axef}). We get:
\begin{flalign}
\nonumber
\psib_0(R\vec{e}_R;a,\beta) & = a \left[ 2+\varpi - \varpi \ln \frac{4 (2+\varpi) \sin \beta}{(2+\varpi) \cos \beta + W} \right.\\
& \qquad \qquad  \qquad \left. -(1+\varpi) \sin 2\beta \ln \left(  1+ \frac{2}{W}\right) \right].
\label{eq:axfloat}
\end{flalign}
For both coherence and optimization, the value of $\psicd$, when necessary (i.e. in the case where $\beta > \frac{\pi}{2}$), must be replaced by its approximation at the same order, namely:
\begin{equation}
\psicd_0(R\vec{e}_R;a) = 2a \left[ \varpi +2 + \varpi \ln \frac{|\varpi|}{4(\varpi+2)} \right].
\label{eq:axcd}
\end{equation}

\begin{figure}
\centering
\includegraphics[width=8.5cm,bb=150 50 554 626,clip==, angle=-90]{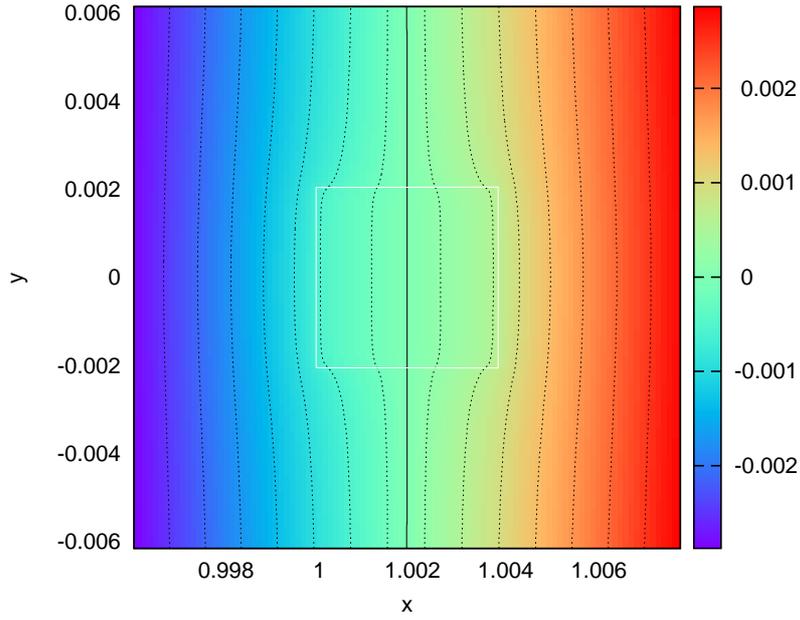}
\caption{Color map showing the relative deviation from Eq.(\ref{eq:deltapsi}) between the exact expression for the potential $\psicell$ and its approximation $\psicell_0$ in the same condition as for Fig. \ref{fig:psidiff.ps}. A few contour levels are superimposed.}
\label{fig:psi_errax.ps}
\end{figure}

The approximation for the potential of the circular sector and then for the polar cell\footnote{Inside the cell, we have $\beta_2 \le \frac{\pi}{2} \le \beta_1$, and so the approximation for the circular sector (i.e. for $\theta \in [\theta_1',\theta_2']$) is:
\begin{flalign}
\psipod(R\vec{e}_R;a,\beta_1,\beta_2) & \approx a \left( 2 \varpi \ln |\varpi|  + \varpi \ln (\sin \beta_1' \sin \beta_2 ) \right.\\ \nonumber
& \left. + \varpi \ln \left\{ \left[(2+\varpi) \cos \beta_2+ W(\beta_2)\right]\left[(2+\varpi) \cos \beta_1'+ W(\beta'_1)\right] \right\} \right.\\ \nonumber
& \left.+ (1+\varpi) \left[ \sin 2\beta_1' \ln \left(  1+ \frac{2}{W(\beta_1')}\right) + \sin 2\beta_2 \ln \left(  1+ \frac{2}{W(\beta_2)}\right)\right] \right)
\label{eq:axpiece}
\end{flalign}
where $\beta_1'=\beta_2 + \frac{\theta_1'+\theta_2'}{2} - \theta = \pi - \beta_1$. The radial acceleration is given by
\begin{equation}
\gr^{\rm sector}(R\vec{e}_R;a,\beta_1,\beta_2) = -\frac{1}{a} \partial_\varpi \psipod(R\vec{e}_R;a,\beta_1,\beta_2),
\end{equation}
and the angular acceleration is
\begin{equation}
\gt^{\rm sector}(R\vec{e}_R;a,\beta_1,\beta_2) = -\frac{1}{R} \partial_\theta \psipod(R\vec{e}_R;a,\beta_1,\beta_2).
\end{equation}
} are then found from Eqs.(\ref{eq:axfloat}) and (\ref{eq:axcd}) by using convenient amplitudes $\beta_1$ and $\beta_2$, and radii $\ain$ and $\aout$. We have checked the accuracy of this approximation in the same conditions as above. The relative difference
\begin{equation}
\Delta \equiv 1-\frac{\psicell_0}{\psicell}
\label{eq:deltapsi}
\end{equation}
is shown in Fig. \ref{fig:psi_errax.ps} for $N=256$ and $L=6N$ inside the cell and close neighbourhood. We note that the agreement is remarkable with $| \Delta| \approx 10 ^{-3}$. It is especially excellent again along the mean arc $R=\frac{1}{2}(\ain+\aout)$, which means that the angular component of the force should be very accurate.

\begin{figure}
\centering
\includegraphics[width=9.5cm,bb=25 46 767 583,clip==]{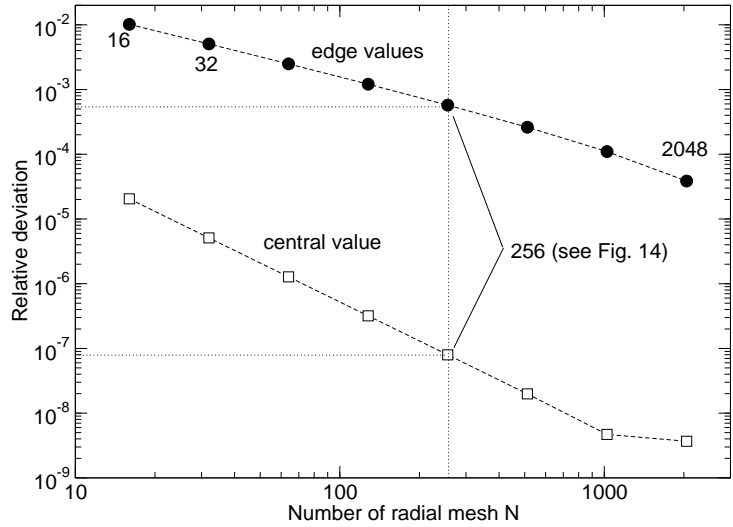}
\caption{Relative deviation $\Delta$ between $\psicell$ given by and its approximation as a function of the resolution parameters $N$ and $L=6N$: at the center of the polar cell ({\it white square}) and along the boundary ({\it filled circles}) (see also Fig.\ref{fig:psi_errax.ps} for $N=256$).}
\label{fig:maxratioax.eps}
\end{figure}

As above, we have changed the resolution (still with $L=6N$). The results are displayed in Fig. \ref{fig:maxratioax.eps}. We roughly find:
\begin{equation}
| \Delta_{\rm boundary} | \approx 5 \times 10^{-4}  \times \left(\frac{256}{N}\right)^{1.1}
\end{equation}
for values of the potential along the boundary, and
\begin{equation}
| \Delta_{\rm c} | \approx 8 \times 10^{-8} \times \left(\frac{256}{N}\right)^{1.8}
\end{equation}
for values at the cell center. 
 
\section{Concluding remarks}

In this article, we have investigated the gravitational potential of the circular sector and polar cell, which are two fundamental patterns present in numerical simulations of disc-like structure. Several new results have been established : the exact formula valid in all space and for any cell shape, an average theorem and an approximation for the interior potential. We have also quantified the curvature effects by comparing the potential of the polar cell and that of the Cartesian cell. All the formulae presented here are appropriate to both theoretical and applied studies requiring either low or high spatial resolution. Applications to electrostatics and capacitance problems are straightforward.

%\begin{acknowledgements}
\bigskip
{\bf Acknowledgements.} I am grateful to the CNU, section 34, for supporting a six-months full-time research project through CRCT-2011 funding delivered by the MESR. It is a pleasure to thank F. Hersant and A. Pierens. 
%\end{acknowledgements}

%\bibliographystyle{unsrt}
\bibliographystyle{spbasic}
%\bibliography{../hurebibtex}

%\bibliographystyle{aa}
%\bibliography{../hurebibtex}

\appendix

\section{Elliptic integrals}
\label{app:ei}

The incomplete elliptic integral of the first kind is
\begin{equation}
F(\phi,k) = \int_0^\phi{\frac{dx}{\sqrt{1-k^2 \sin^2 x}}},
\end{equation}
where $k$ is the modulus and $\phi$ the amplitude. The incomplete elliptic integral of the second kind is
\begin{equation}
E(\phi,k) = \int_0^\phi{dx \sqrt{1-k^2 \sin^2 x}},
\end{equation}
The incomplete elliptic integral of the third kind is
\begin{equation}
\Pi(\phi,m^2,k) = \int_0^\phi{\frac{dx}{(1-m^2 \sin^2 x) \sqrt{1-k^2 \sin^2 x}}},
\end{equation}
where $m$ is the characteristic or parameter. The complete elliptic integrals are found by setting $\phi=\frac{\pi}{2}$, namely
\begin{equation}
F(\frac{\pi}{2},k) \equiv \elik(k),
\end{equation}
\begin{equation}
E(\frac{\pi}{2},k) \equiv \elie(k),
\end{equation}
and
\begin{equation}
\Pi(\frac{\pi}{2},m^2,k) \equiv \elipi(m^2,k).
\end{equation}

\section{Acceleration of the circular disc}
\label{app:g}

The two components of the acceleration $\vec{g}^{\rm \, disc}=-\nabla \psicd$ associated with Eq.(\ref{eq:psi_as}) are also given in Durand's textbook \citep{durand64}. Up to a factor $-G \Sigma$, these are:
\begin{equation}
-\pdr \psicd=2 \frac{\dd}{R}\left[ \left(1- \frac{k^2}{2} \right) \elik(k) -  \elie(k)  \right],
\end{equation}
and
\begin{equation}
-\pdz \psicd = 2 \pi\sign(\zeta) \epsilon' + \frac{2\zeta}{\dd} \left[\frac{R-a}{R+a} \elipi(m,k) - \elik(k)\right],
\end{equation}
where $\sign(\zeta)$  is the sign of $\zeta$ (or $0$ if $\zeta=0$).

\section{Derivative of the elliptic integrals}
\label{app:dei}

For the incomplete elliptic integrals of the first and second kinds, we have respectively for any modulus $k$, complementary modulus $k'=\sqrt{1-k^2}$ and amplitude $\phi$ \citep{gradryz07}
\begin{equation}
k {k'}^2 \partial_k F(\phi,k) = E(\phi,k)-{k'}^2 F(\phi,k)- \frac{k^2 \sin \phi \cos \phi}{\sqrt{1- k^2 \sin^2 \phi} },
\label{eq:partialk}
\end{equation}
and
\begin{equation}
k \partial_k E(\phi,k) = E(\phi,k)-F(\phi,k).
\label{eq:partiale}
\end{equation}
Regarding the incomplete elliptic integrals of the third kind, we have two derivatives. We have
\begin{equation}
\partial_k \Pi(\phi,m^2,k) = k \int_0^\phi{\frac{\sin^2 \phi d \phi}{(1-m^2 \sin^2 \phi) \sqrt{1-k^2 \sin^2 \phi}^3}},
\end{equation}
and so
\begin{flalign}
\frac{k^2-m^2}{k} \partial_k \Pi(\phi,m^2,k) & = \int_0^\phi{\frac{(k^2-m^2)\sin^2 \phi d \phi}{(1-m^2 \sin^2 \phi) \sqrt{1-k^2 \sin^2 \phi}^2}}\\ \nonumber
% & = \int_0^\phi{\frac{(k^2 \sin^2 \phi -1 -m^2 \sin^2 \phi +1)d \phi}{(1-m^2 \sin^2 \phi) \sqrt{1-k^2 \sin^2 \phi}}}\\ \nonumber
 & = - \Pi(\phi,m^2,k)+  \int_0^\phi{\frac{d \phi}{\sqrt{1-k^2 \sin^2 \phi}^3}}\\ \nonumber
 & = - \Pi(\phi,m^2,k)+ \frac{E(\phi,k)}{{k'}^2} - \frac{k^2 \sin \phi \cos \phi}{{k'}^2\sqrt{1-k^2 \sin^2 \phi}}
\end{flalign}
where the last integral can be found in \cite{gradryz07}. The partial derivative with respect to $k$ is then given by
\begin{flalign}
\partial_k \Pi(\phi,m^2,k) & = \frac{k}{m^2-k^2} \left[\Pi(\phi,m^2,k) - \frac{E(\phi,k)}{{k'}^2} + \frac{k^2 \sin \phi \cos \phi}{{k'}^2\sqrt{1-k^2 \sin^2 \phi}} \right],
\end{flalign}
which is the incomplete version of the expression in \cite{durand64}. The partial derivative with respect to the parameter $m$ is:
\begin{equation}
\frac{1}{2m}\partial_m \Pi(\phi,m^2,k) = \int_0^\phi{\frac{\sin^2 \phi d \phi}{(1-m^2 \sin^2 \phi)^2 \sqrt{1-k^2 \sin^2 \phi}}}.
\end{equation}

The integral (which can be obtained by an arduous calculation from the elementary decomposition of the denominator) is also given by Dieckmann (2011)\footnote{See \tt http://pi.physik.uni-bonn.de/\~{}dieckman/}. We then have:
\begin{flalign}
m{m'}^2\partial_m \Pi(\phi,m^2,k) & = \frac{m^2}{m^2-k^2} E(\phi,k) - F(\phi,k)  +\frac{m^4-k^2}{m^2-k^2} \Pi(\phi,m^2,k) \\ \nonumber
&  \qquad  - \frac{m^4 \sin \phi \cos \phi \sqrt{1-k^2 \sin^2 \phi}}{(m^2-k^2)(1-m^2 \sin^2 \phi)},
\end{flalign}
where  $m'=\sqrt{1-m^2}$ is the parameter complementary of $m$.

\section{Partial derivatives with respect to the radius $a$}
\label{app:3der}

The partial derivative of the first term in Eq.(\ref{eq:3der}) is
\begin{flalign}
\partial_a \left[ \dd E(\phi,k) \right] & = \frac{a+R}{\dd} E(\phi,k) + \dd \partial_a k \times \partial_k E(\phi,k)\\ \nonumber
& = \frac{1}{\dd}\left(a+R + \frac{\zeta^2+R^2-a^2}{2a} \right) E(\phi,k) - \frac{\zeta^2+R^2-a^2}{2a\dd} F(\phi,k).
\end{flalign}
where $\delta$, $k$ and $m$ are given by Eqs.(\ref{eq:r1}), (\ref{eq:kmod}) and (\ref{eq:mmod}) respectively. For the second term, we have
\begin{flalign}
\nonumber
\partial_a \left[ \frac{a^2-R^2}{\dd}F(\phi,k) \right]  = \frac{2a\dd^2-(a^2-R^2)(a+R)}{\dd^3} F(\phi,k) +  \frac{a^2-R^2}{\dd} \partial_a k \times \partial_k F(\phi,k)\\
\nonumber
 = \frac{(a^2-R^2)(\zeta^2+R^2-a^2)}{2a\dd^3}  \frac{E(\phi,k)}{{k'}^2}\\ \nonumber
 \quad +\frac{1}{\dd^3}\left[ 2a\zeta^2+(a+R)^3 - \frac{(a^2-R^2)(\zeta^2+R^2-a^2)}{2a}  \right] F(\phi,k)\\
 \quad - \frac{(a^2-R^2)(\zeta^2+R^2-a^2)}{2a\dd^3}\frac{k^2 \sin \phi \cos \phi}{{k'}^2\sqrt{1- k^2 \sin^2 \phi} }
\end{flalign}
The last term containing the elliptic integral of the third kind is not easy to determine, since both $m$ and $k$ depend on $a$. We have
\begin{equation}
\partial_a \Pi(\phi,m^2,k) = \partial_k  \Pi(\phi,m^2,k) \times \partial_a k + \partial_m  \Pi(\phi,m^2,k) \times \partial_a m 
\end{equation}
and so
\begin{flalign}
\partial_a \Pi(\phi,m^2,k)  % = \partial_k  \Pi(\phi,m^2,k) \times \partial_a k + \partial_m  \Pi(\phi,m^2,k) \times \partial_a m \\ \nonumber
 =   \left[\Pi(\phi,m^2,k) - \frac{E(\phi,k)}{{k'}^2} + \frac{k^2 \sin \phi \cos \phi}{{k'}^2\sqrt{1-k^2 \sin^2 \phi}} \right] \frac{k}{m^2-k^2}  \partial_a k  \\\nonumber
 + \left[  \frac{m^2}{m^2-k^2} E(\phi,k) - F(\phi,k) +\frac{m^4-k^2}{m^2-k^2} \Pi(\phi,m^2,k) \right. \\\nonumber
 -  \left.  \frac{m^4 \sin \phi \cos \phi \sqrt{1-k^2 \sin^2 \phi}}{(m^2-k^2)(1-m^2 \sin^2 \phi)} \right] \frac{1}{m{m'}^2} \partial_a m 
\end{flalign}
We also have
\begin{equation}
\frac{k}{m^2-k^2} \partial_a k = \frac{(a+R)^2(R^2+\zeta^2-a^2)}{2a \dd^2 \zeta^2} = \frac{R^2+\zeta^2-a^2}{\zeta^2} \frac{k^2}{2am^2},
\end{equation}
and
\begin{equation}
\frac{m}{m^2-k^2}\frac{1}{{m'}^2} \partial_a m = \frac{(R-a)\dd^2}{2(R+a)a{m'}^2\zeta^2} = \frac{\dd^2}{\zeta^2} \frac{1}{2a} \frac{R+a}{R-a},
\end{equation}
where
\begin{equation}
m^4-k^2 = m^2(1-{m'}^2)-k^2 = m^2 \left( \frac{\zeta^2}{\dd^2}-{m'}^2 \right).
\end{equation}
After some algebra, we find
\begin{flalign}
\partial_a \Pi(\phi,m^2,k)  = -&\frac{a+R}{a-R} \frac{1}{\dd^2}\left[3R-a+\frac{2R \zeta^2}{(a+R)^2} \right]  \Pi(\phi,m^2,k)  \\ \nonumber
  +& \frac{a-R}{2a(a+R)} \frac{1}{{m'}^2} F(\phi,k)  -\frac{a+R}{2(a-R)a} \frac{R^2+3a^2+\zeta^2}{(a-R)^2+\zeta^2} E(\phi,k)\\ \nonumber
  + &\frac{m^2}{2a\zeta^2} \sin \phi \cos \phi \left[ \frac{k^4}{{k'}^2 m^4} \frac{R^2+\zeta^2-a^2}{\sqrt{1-k^2 \sin^2 \phi}} + \frac{\sqrt{1-k^2 \sin^2 \phi}}{(1-m^2 \sin^2 \phi)} \frac{a+R}{a-R} \dd^2 \right].
\end{flalign}
As we have
\begin{equation}
\partial_a \left(\frac{a-R}{a+R}\frac{1}{\dd} \right) = \frac{1}{\dd^3}\left[3R-a+\frac{2R \zeta^2}{(a+R)^2} \right],
\end{equation}
we deduce that the partial derivative of the third term in Eq.(\ref{eq:3der}) does not depend on $\Pi$. Actually, we get
\begin{flalign}
\partial_a  \left[\frac{a-R}{a+R}\frac{\zeta^2}{\dd} \Pi(\phi,m^2,k) \right]=
& \frac{\zeta^2}{2a\dd} F(\phi,k) -\frac{\zeta^2}{2a\dd} \frac{R^2+3a^2+\zeta^2}{(a-R)^2+\zeta^2} E(\phi,k)\\ \nonumber
& + \frac{1}{2a\dd} \frac{a-R}{a+R} \frac{\sin \phi \cos \phi}{\sqrt{1-k^2 \sin^2 \phi}}\frac{k^4}{m^2 {k'}^2}(R^2+\zeta^2-a^2)\\ \nonumber
& + \frac{m^2 \dd}{2a} \frac{\sin \phi \cos \phi \sqrt{1-k^2 \sin^2 \phi}}{1-m^2 \sin^2 \phi}.
\end{flalign}

\section{Integration of the asymmetric term}
\label{app:intasymetric}

We first separate the term into two terms, namely
\begin{flalign}
\frac{\sqrt{1-k^2 \sin^2 \phi}}{(1-m^2 \sin^2 \phi)} & = \frac{1}{\sqrt{1-k^2 \sin^2 \phi}} \left( \frac{1-k^2 \sin^2 \phi}{(1-m^2 \sin^2 \phi)} \right)\\ \nonumber
& =  \frac{1}{\sqrt{1-k^2 \sin^2 \phi}} \left( \frac{1}{1-m^2 \sin^2 \phi}-\frac{k^2}{m^2} \frac{m^2\sin^2 \phi}{1-m^2 \sin^2 \phi} \right)\\ \nonumber
& =  \frac{1}{\sqrt{1-k^2 \sin^2 \phi}} \left[ \frac{1}{1-m^2 \sin^2 \phi} \left(1-\frac{k^2}{m^2} \right) + \frac{k^2}{m^2} \right].
\end{flalign}
We then get
\begin{flalign}
\nonumber
\frac{m^2 \dd}{2a} \sin \phi \cos \phi  \frac{\sqrt{1-k^2 \sin^2 \phi}}{(1-m^2 \sin^2 \phi)} & = \frac{2R \dd}{(a+R)^2}  \frac{\sin \phi \cos \phi}{\sqrt{1-k^2 \sin^2 \phi}} \\\nonumber
& \qquad \times \left[\frac{k^2}{m^2} + \frac{1}{1-m^2 \sin^2 \phi} \left(1-\frac{k^2}{m^2} \right)  \right]\\ \nonumber
&  =  \frac{2R \sin \phi \cos \phi}{\sqrt{\dd^2 - 4aR \sin^2 \phi}}\\
&  + \frac{2R \zeta^2  \sin \phi \cos \phi}{[(a+R)^2-4aR \sin^2 \phi]\sqrt{\dd^2-4aR\sin^2 \phi}},
\end{flalign}
where $\dd^2-4aR\sin^2 \phi = $PP$'^2$. The integration of the first term (if we omit the quantity $2R \sin \phi \cos \phi$) with respect to $a$ is:
\begin{equation}
\int_a{\frac{da}{\sqrt{\dd^2 - 4aR \sin^2 \phi}}} = \asinh \frac{a+R \cos 2 \phi}{\sqrt{\zeta^2+R^2 \sin^2 2 \phi}} .
\end{equation}
Regarding the second term, we set
%\begin{equation}
%t = a+R \cos 2 \phi,
%\end{equation}
%and so
%\begin{equation}
%\int_a{\frac{da}{[(a+R)^2-4aR \sin^2 \phi]\sqrt{(a+R)^2+\zeta^2-4aR\sin^2 \phi}}} = \int{\frac{dt}{(t^2+R^2 \sin^2 2 \phi)\sqrt{t^2+\zeta^2+R^2 \sin^2 2 \phi}}}
%\end{equation}
%We then set
\begin{equation}
%u = \frac{t}{\sqrt{t^2+\zeta^2+R^2 \sin^2 2 \phi}}.
u = \frac{a+R \cos 2 \phi}{\sqrt{(a+R \cos 2 \phi)^2+\zeta^2+R^2 \sin^2 2 \phi}} =  \frac{a+R \cos 2 \phi}{{\rm PP}'}.
\end{equation}
As
\begin{equation}
%\frac{du}{1-u^2} = \frac{dt}{\sqrt{t^2+\zeta^2+R^2 \sin^2 2 \phi}},
\frac{du}{1-u^2} = \frac{da}{\sqrt{(a+R \cos 2 \phi)^2+\zeta^2+R^2 \sin^2 2 \phi}} = \frac{da}{{\rm PP}'},
\end{equation}
the integral of this term now reads (if we omit the quantity $2 \zeta^2 R \sin \phi \cos \phi$):
\begin{flalign}
\int_a{\frac{da}{[(a+R)^2-4aR \sin^2 \phi]{\rm PP}'}} & = \int{\frac{du}{\zeta^2 u^2 + R^2 \sin^2 2 \phi}} \\\nonumber
%& = \frac{1}{|\zeta R \sin 2\phi |}\int{\frac{d\left| \frac{\zeta}{R \sin 2\phi} \right| u}{\frac{\zeta^2  u^2}{ R^2 \sin^2 2 \phi} +1}} \\\nonumber
 & = \frac{1}{|\zeta R \sin 2\phi |} \atan \left[ \left| \frac{\zeta}{R \sin 2\phi} \right|\frac{a+R\cos 2 \phi}{{\rm PP}'} \right] \\\nonumber
 & = \frac{1}{\zeta R \sin 2\phi } \atan \left[ \frac{\zeta(a+R\cos 2 \phi)}{R \sin 2\phi {\rm PP}'} \right].
\end{flalign}

We can then write the second term in the right-hand-side of Eq.(\ref{eq:almostfinal}) as a derivative, like
\begin{flalign}
\frac{m^2 \dd}{2a} \frac{\sin \phi \cos \phi \sqrt{1-k^2 \sin^2 \phi}}{(1-m^2 \sin^2 \phi)} & = \partial_a \left\{ R \sin 2 \phi 
\asinh \frac{a+R \cos 2 \phi}{\sqrt{\zeta^2+R^2 \sin^2 2 \phi}} \right. \\\nonumber
& \qquad \qquad \left.  + \zeta \atan \left[ \frac{\zeta(a+R\cos 2 \phi)}{R \sin 2\phi {\rm PP}'} \right]\right\}.
\end{flalign}

\section{Radial acceleration of the floating sector}
\label{app:grpod}

To determine the radial component of the acceleration $\vec{g}^{\rm float}=-\vec{\nabla} \psib$ associated with Eq.(\ref{eq:psifloatfinal}), we need the partial derivatives of $k$ and $m$ with respect to $R$. We have
\begin{equation}
\frac{1}{k}\partial_R k = \frac{\zeta^2+a^2-R^2}{2R\dd^2},
\label{eq:partialkr}
\end{equation}
and
\begin{equation}
\frac{1}{m}\partial_R m = \frac{a-R}{2R(a+R)}.
\label{eq:partialmr}
\end{equation}
These relations can be deduced from Eqs.(\ref{eq:partialka}) and (\ref{eq:partialma}) respectively by exchanging $a$ and $R$. With the help of the Appendix \ref{app:3der}, we find
\begin{flalign}
\partial_R \left[ \dd E(\phi,k) \right] &  = \frac{a+R}{\dd} E(\phi,k) + \dd \partial_R k \times \partial_k E(\phi,k)\\ \nonumber
& = \frac{1}{\dd}\left(a+R + \frac{\zeta^2+a^2-R^2}{2R} \right) E(\phi,k) - \frac{\zeta^2+a^2-R^2}{2R\dd} F(\phi,k),
\end{flalign}
and
\begin{flalign}
\partial_R \left[ \frac{a^2-R^2}{\dd}F(\phi,k) \right] & = \frac{-2R\dd^2-(a^2-R^2)(a+R)}{\dd^3} F(\phi,k) +  \frac{a^2-R^2}{\dd} \partial_R k \times \partial_k F(\phi,k)\\
\nonumber
& = \frac{(a^2-R^2)(\zeta^2+a^2-R^2)}{2R\dd^3}  \frac{E(\phi,k)}{{k'}^2}\\ \nonumber
& \quad -\frac{1}{\dd^3}\left[ 2R\zeta^2 +(a+R)^3 + \frac{(a^2-R^2)(\zeta^2+a^2-R^2)}{2R}  \right] F(\phi,k)\\
& \quad - \frac{(a^2-R^2)(\zeta^2+a^2-R^2)}{2R\dd^3}\frac{k^2 \sin \phi \cos \phi}{{k'}^2\sqrt{1- k^2 \sin^2 \phi} },
\end{flalign}
where $k'=\sqrt{1-k^2}$, and
\begin{flalign}
\partial_R  \left[\frac{a-R}{a+R}\frac{\zeta^2}{\dd} \Pi(\phi,m,k) \right] &=  - \frac{\zeta^2}{2R\dd} F(\phi,k) +\frac{\zeta^2}{2R\dd} \frac{a^2+3R^2+\zeta^2}{(a-R)^2+\zeta^2} E(\phi,k)\\ \nonumber
& + \frac{1}{2R\dd} \frac{a-R}{a+R} \frac{\sin \phi \cos \phi}{\sqrt{1-k^2 \sin^2 \phi}}\frac{k^4}{m^2 {k'}^2}(a^2+\zeta^2-R^2)\\ \nonumber
& - \frac{m^2 \dd}{2R} \frac{\sin \phi \cos \phi \sqrt{1-k^2 \sin^2 \phi}}{1-m^2 \sin^2 \phi}
\end{flalign}

For the derivative of the two functions $RH$ and $\zeta T$, we respectively find
\begin{flalign}
\pdr RH(\phi)= H(\phi)-\frac{R\sin 2 \phi}{\sqrt{\zeta^2+R^2 \sin^2 2 \phi}} \left( \frac{\zeta^2 \cos 2 \phi -aR \sin^2 2 \phi}{{\rm PP}'}- \frac{\zeta^2 \cos 2 \phi}{\sqrt{\zeta^2+R^2}} \right)
\end{flalign}
and
\begin{flalign}
\pdr \zeta T(\phi) & =- \zeta^2 \sin 2 \phi \frac{ {\rm PP}'}{R^2 \sin^2 2 \phi {{\rm PP}'}^2+\zeta^2(a+R \cos 2 \phi)^2}\\ \nonumber
& \qquad \qquad \times \left[ R \cos 2 \phi - (a+R \cos 2 \phi)\left(1+\frac{R(R+a \cos 2 \phi)}{{{\rm PP}'}^2}\right) \right]\\ \nonumber
& + \zeta^2 \sin 2 \phi \frac{R\cos 2\phi}{(\zeta^2+R^2 \sin^2 2\phi)\sqrt{\zeta^2+R^2}}
\end{flalign}

So the radial gradient of $\psib$ is
\begin{flalign}
\pdr \psib = \frac{\dd}{R}\left[ E(\phi,k)- \left(1- \frac{k^2}{2} \right) F(\phi,k) - \frac{m^2}{2} \frac{\sin \phi \cos \phi \sqrt{1-k^2 \sin^2 \phi}}{1-m^2 \sin^2 \phi} \right]\\ \nonumber
 + H(\phi)-\frac{R\sin 2 \phi}{\zeta^2+R^2 \sin^2 2 \phi} \left( \frac{\zeta^2 \cos 2 \phi -aR \sin^2 2 \phi}{{\rm PP}'} - \frac{\zeta^2 \cos 2 \phi}{\sqrt{\zeta^2+R^2}} \right)\\ \nonumber
- \zeta^2 \sin 2 \phi \frac{ {\rm PP}'}{R^2 \sin^2 2 \phi {{\rm PP}'}^2+\zeta^2(a+R \cos 2 \phi)^2}\\ \nonumber
\times \left[ R \cos 2 \phi - (a+R \cos 2 \phi)\left(1+\frac{R(R+a \cos 2 \phi)}{{{\rm PP}'}^2}\right) \right]\\ \nonumber
+ \zeta^2 \sin 2 \phi \frac{R \cos 2 \phi }{(\zeta^2+R^2\sin^2 2 \phi) \sqrt{R^2+\zeta^2}}.
\end{flalign}

We see that this expression is fully compatible with the expression for $\pdr \psicd$ given in the Appendix \ref{app:g} for $\phi=\frac{\pi}{2}$, since we get in this case
\begin{flalign}
\pdr \psib & = \frac{\dd}{R}\left[ \elie(k)- \left(1- \frac{k^2}{2} \right) \elik(k) \right]\\ \nonumber
& = \frac{1}{2}\pdr \psicd.
\end{flalign}

\section{Angular acceleration of the floating sector}
\label{app:gthetapod}

Regarding the angular acceleration $\partial_\theta \psib$ of the floating sector, we have
\begin{equation}
\partial_\theta = \frac{1}{2}\partial_\beta,
\end{equation}
meaning that we need in particular the partial derivatives of the elliptic integrals with respect to their amplitude $\beta$. For any amplitude $\phi$, we have
\begin{equation}
\partial_\phi F(\phi,k) = \frac{1}{\sqrt{1-k^2 \sin^2 \phi}},
\end{equation}
\begin{equation}
\partial_\phi E(\phi,k) = \sqrt{1-k^2 \sin^2 \phi},
\end{equation}
and
\begin{equation}
\partial_\phi \Pi(\phi,m,k) = \frac{1}{ (1-m^2 \sin^2 \phi)\sqrt{1-k^2 \sin^2 \phi}}.
\end{equation}
None of the quantities $k$, $\delta$, $k$ and $m$ depend on $\phi$. The inverse hyperbolic term and the inverse trigonometric term in Eq.(\ref{eq:psipodbasic}) give respectively
\begin{flalign}
\partial_\phi  \left( \sin 2\phi \asinh \frac{a+R \cos 2 \phi}{\sqrt{\zeta^2+R^2 \sin^2 2 \phi}} \right ) & = 2 \cos 2\phi \asinh \frac{a+R \cos 2 \phi}{\sqrt{\zeta^2+R^2 \sin^2 2 \phi}} \\\nonumber
& \qquad - 2 \sin^2 2\phi  \frac{R(\zeta^2 +R^2+aR\cos 2 \phi)}{{\rm PP}'(\zeta^2+R^2 \sin^2 2 \phi)}
\end{flalign}
and
\begin{flalign}
\nonumber
\partial_\phi \left\{ \zeta \atan \left[ \frac{\zeta(a+R\cos 2 \phi)}{R \sin 2\phi \; {\rm PP}'} \right] \right\} = -\frac{2 \zeta^2 R \; {\rm PP}' }{(a+R)^2(1-m^2\sin^2 \phi)(\zeta^2+R^2 \sin^2 2 \phi)}\\
\times \left[ R+a\cos 2\phi - (a+R \cos 2\phi) k^2 \frac{\sin^2\phi \cos^2 \phi}{1-k^2\sin^2\phi} \right]
\end{flalign}
Finally, we have
\begin{flalign}
\pdtheta \psib & = \frac{{\rm PP}'}{2} + \frac{a^2-R^2}{2{\rm PP}'} + \frac{a-R}{a+R} \frac{\zeta^2}{2 {\rm PP}'(1-m^2 \sin^2 \phi)}\\ \nonumber
& + \cotan 2 \phi \; H(\phi)-\sin^2 2\phi \left[ \frac{R(\zeta^2 +R^2+aR\cos 2 \phi)}{{\rm PP}'(\zeta^2+R^2 \sin^2 2 \phi)} - \frac{R \sqrt{R^2+\zeta^2}}{\zeta^2+R^2 \sin^2 2 \phi} \right] \\ \nonumber
& -\frac{\zeta^2}{(\zeta^2+R^2 \sin^2 2 \phi)} \left\{  \frac{R{\rm PP}'}{(a+R)^2(1-m^2\sin^2 \phi)} \left[ R+a\cos 2\phi - (a+R \cos 2\phi) k^2 \frac{\sin^2\phi \cos^2 \phi}{1-k^2\sin^2\phi} \right] \right. \\\nonumber
&\left. \qquad\qquad\qquad\qquad\qquad\qquad- \sqrt{R^2+\zeta^2} \right\}
\end{flalign}

\section{Vertical acceleration of the floating sector}
\label{app:gzpod}

We have $\partial_Z = \partial_\zeta$ and so
\begin{flalign}
\partial_\zeta \delta E(\phi,k) = \frac{\zeta F(\phi,k)}{\delta},
\end{flalign}
\begin{flalign}
\partial_\zeta \frac{a^2-R^2}{\delta}F(\phi,k) = - \frac{\zeta(a^2-R^2)}{\delta^3} \left[ \frac{E(\phi,k)}{{k'}^2} - \frac{k^2\sin \phi \cos \phi}{{k'}^2\sqrt{1-k^2 \sin^2 \phi}} \right],
\end{flalign}
where $k'=\sqrt{1-k^2}$, and
\begin{flalign}
\partial_\zeta \frac{a-R}{a+R}\frac{\zeta^2}{\delta}\Pi(\phi,m^2,k) & = \frac{a-R}{a+R} \frac{\zeta}{\delta} \left[ \Pi(\phi,m^2,k)  + \frac{k^2}{m^2 {k'}^2} E(\phi,k) \right.\\\nonumber
& \qquad \qquad  \qquad \qquad  \left. -  \frac{k^4\sin \phi \cos \phi}{m^2{k'}^2\sqrt{1-k^2 \sin^2 \phi}} \right].
\end{flalign}
The trigonometric term gives
\begin{flalign}
\partial_\zeta \zeta T(\phi) & = T(\phi) - \zeta \sin 2\phi \frac{R ({{\rm PP}'}^2-\zeta^2)(a+R\cos 2\phi)}{R^2 \sin^2 2 \phi {{\rm PP}'}^2+\zeta^2(a+R \cos 2 \phi)^2},\\\nonumber
&  = T(\phi) - \sin 2\phi  \frac{\zeta}{{\rm PP}'} \frac{R (a+R\cos 2\phi)}{R^2 \sin^2 2 \phi +\zeta^2},
\end{flalign}
and the hyperbolic term gives
\begin{equation}
\partial_\zeta RH(\phi) = \frac{R\zeta \sin 2\phi}{\zeta^2+R^2 \sin^2 2\phi} \left( \frac{a+R\cos 2\phi}{{\rm PP}'}- \frac{R\cos 2\phi}{\sqrt{\zeta^2+ R^2 \sin^2 2 \phi}}\right),
\end{equation}
Gathering terms, we finally find
\begin{equation}
\partial_\zeta \psib  =\frac{\zeta}{\delta} \left[ F(\phi,k) -  \frac{R-a}{a+R} \Pi(\phi,m^2,k) \right] + T(\phi).
\end{equation}
This expression is also compatible with the expression for $\partial_\zeta \psicd$ given in the Appendix \ref{app:g} for $\phi=\frac{\pi}{2}$ since the term $T(\phi)$ is $-|\zeta|\pi$, $-|\zeta|\frac{\pi}{2}$ or $0$ depending on the ratio $R/a$, according to Eq.(\ref{eq:tatpiover2}).

\section{Potential of a rectangular sheet}
\label{app:rectplate}

The potential in the plane of a rectangular sheet (up to a factor $-G\Sigma$) is given by the expression \citep{durand64,cha12}:
\begin{flalign}
\label{eq:psi_rectplate}
\psirect(\vec{r};x'_1,x'_2,y'_1,y'_2)  = (y'_2-y) \ln \frac{x_2'-x + \sqrt{(x_2'-x)^2+(y'_2-y)^2}}{x_1'-x + \sqrt{(x_1'-x)^2+(y'_2-y)^2}}\\
\nonumber
 -(y'_1-y) \ln \frac{x_2'-x + \sqrt{(x_2'-x)^2+(y'_1-y)^2}}{x_1'-x + \sqrt{(x_1'-x)^2+(y'_1-y)^2}}\\
\nonumber
+ (x'_2-x) \ln \frac{y_2'-y + \sqrt{(x_2'-x)^2+(y'_2-y)^2}}{y_1'-y + \sqrt{(x_2'-x)^2+(y'_1-y)^2}}\\
\nonumber
 -(x'_1-x) \ln \frac{y_2'-y + \sqrt{(x_1'-x)^2+(y'_2-y)^2}}{y_1'-y + \sqrt{(x_1'-x)^2+(y'_1-y)^2}},
\end{flalign}
where $\vec{r}=x\vec{e}_x+y\vec{e}_y$, and $(x'_1,y'_1)$, $(x'_2 \ge x'_1, y'_1)$, $(x'_2, y'_2\ge y'_1)$ and $(x_1',y_2')$ denote the Cartesian coordinates of the four corners of the sheet.

\end{document}